\documentclass[acmsmall]{acmart}
\usepackage{comment}
\usepackage{booktabs}
\usepackage{arydshln} 
\usepackage{graphicx}
\usepackage{xspace}
\usepackage{hyperref}
\usepackage{esvect}
\usepackage{enumitem}
\usepackage{wrapfig}
\setitemize{noitemsep,topsep=0pt,parsep=0pt,partopsep=0pt}
\AtBeginDocument{%
  \providecommand\BibTeX{{%
    \normalfont B\kern-0.5em{\scshape i\kern-0.25em b}\kern-0.8em\TeX}}}

\setcopyright{acmcopyright}
\copyrightyear{2021}
\acmYear{2021}
\acmDOI{10.1145/1122445.1122456}

\acmJournal{TCPS}
\acmVolume{5}
\acmNumber{3}
\acmArticle{111}
\acmMonth{7}

\begin{document}
\makeatletter
\newcommand\CO[1]{%
  \@tempdima=\linewidth%
  \advance\@tempdima by -2\fboxsep%
  \advance\@tempdima by -2\fboxrule%
  \leavevmode\par\noindent%
  \fbox{\parbox{\the\@tempdima}{%
    \small\normalfont\sffamily #1}}%
  \smallskip\par}
\makeatother

\newcommand{\red}[1]{\textcolor{red}{#1} }

\newcommand{\code}[1]{\texttt{#1}}

\newcommand{\eg}{{e.g.}}
\newcommand{\ie}{{i.e.}}
\newcommand{\etal}{\emph{et al.}}
\newcommand{\cf}{{\em cf.}}

\newcommand\tmr{\texttt{TMR}\xspace}
\newcommand\hmd{\texttt{HmD}\xspace}
\newcommand\htd{\texttt{HtD}\xspace}
\newcommand\safmon{\texttt{MonAct}\xspace}
\newcommand\sa{\texttt{SA}\xspace}
\newcommand\watchdog{\texttt{WD}\xspace}
\newcommand\sanity{\texttt{SanChk}\xspace}
\newcommand\nprog{\texttt{NProg}\xspace}

\newcommand\machineryOld{\textsc{SafPat}\xspace}
\newcommand\machinery{\textsc{SafSecPat}\xspace}

\newcommand\tool{\textsc{Pattern Synthesis}\xspace}

\newcommand\before{\ensuremath{\mathsf{before}}}
\newcommand\dsl{\ensuremath{\mathsf{SafPat}}}
\newcommand\cp{\ensuremath{\mathsf{cp}}}
\newcommand\subcp{\ensuremath{\mathsf{subcp}}}
\newcommand\ch{\ensuremath{\mathsf{ch}}}
\newcommand\info{\ensuremath{\mathsf{if}}}
\newcommand\hw{\ensuremath{\mathsf{hw}}}
\newcommand\sw{\ensuremath{\mathsf{sw}}}
\newcommand\dep{\ensuremath{\mathsf{dep}}}
\newcommand\id{\ensuremath{\mathsf{id}}}
\newcommand\interface{\ensuremath{\mathsf{interface}}}
\newcommand\ecu{\ensuremath{\mathsf{ecu}}}
\newcommand\Dep{\ensuremath{\mathsf{dep}}}
\newcommand\hz{\ensuremath{\mathsf{hz}}}
\newcommand\fl{\ensuremath{\mathsf{fl}}}
\newcommand\ft{\ensuremath{\mathsf{ft}}}
\newcommand\dm{\ensuremath{\mathsf{dm}}}
\newcommand\flToHz{\ensuremath{\mathsf{fl2Hz}}}
\newcommand\ftTofl{\ensuremath{\mathsf{ft2fl}}}
\newcommand\tp{\ensuremath{\mathsf{fl_{tp}}}}
\newcommand\sev{\ensuremath{\mathsf{pt_{sv}}}}
\newcommand\trsev{\ensuremath{\mathsf{th_{sv}}}}
\newcommand\subhz{\ensuremath{\mathsf{subHz}}}
\newcommand\err{\ensuremath{\mathsf{err}}}
\newcommand\loss{\ensuremath{\mathsf{loss}}}
\newcommand\omission{\ensuremath{\mathsf{omission}}}
\newcommand\late{\ensuremath{\mathsf{late}}}
\newcommand\early{\ensuremath{\mathsf{early}}}
\newcommand\minor{\ensuremath{\mathsf{min}}}
\newcommand\major{\ensuremath{\mathsf{maj}}}
\newcommand\hazardeous{\ensuremath{\mathsf{haz}}}
\newcommand\cat{\ensuremath{\mathsf{cat}}}
\newcommand\safMon{\ensuremath{\mathsf{safMon}}}
\newcommand\sm{\ensuremath{\mathsf{sm}}}
\newcommand\watchDog{\ensuremath{\mathsf{watchDog}}}
\newcommand\wdcp{\ensuremath{\mathsf{wd}}}
\newcommand\nProg{\ensuremath{\mathsf{nProg}}\xspace}
\newcommand\dProg{\ensuremath{\mathsf{2Prog}}\xspace}
\newcommand\voter{\ensuremath{\mathsf{Voter}}}
\newcommand\VOTER{\ensuremath{\mathsf{VT}}}
\newcommand\fs{\ensuremath{\mathsf{fs}}}
\newcommand\candsl{\ensuremath{\mathsf{can}}}

\newcommand\sg{\ensuremath{\mathsf{sg}}}
\newcommand\fop{\ensuremath{\mathsf{f_{op}}}}
\newcommand\fsilent{\ensuremath{\mathsf{f_{sl}}}}
\newcommand\fsafe{\ensuremath{\mathsf{f_{sf}}}}

\newcommand\allNFail{\ensuremath{\mathsf{allXfail}}}
\newcommand\mostNFail{\ensuremath{\mathsf{mostXfail}}}
\newcommand\never{\ensuremath{\mathsf{never}}}

\newcommand\idd{\ensuremath{\mathsf{\id_{d}}}}
\newcommand\idfl{\ensuremath{\mathsf{\id_{fl}}}}
\newcommand\idft{\ensuremath{\mathsf{\id_{ft}}}}
\newcommand\idhz{\ensuremath{\mathsf{\id_{hz}}}}
\newcommand\system{\ensuremath{\mathsf{sys}}}
\newcommand\hzsev{\ensuremath{\mathsf{hz_{sev}}}}
\newcommand\hzexp{\ensuremath{\mathsf{hz_{exp}}}}
\newcommand\hzctl{\ensuremath{\mathsf{hz_{ctl}}}}
\newcommand\idc{\ensuremath{\mathsf{\id_{cp}}}}
\newcommand\idhw{\ensuremath{\mathsf{\id_{hw}}}}
\newcommand\idcOne{\ensuremath{\mathsf{\id_{c1}}}}
\newcommand\idcTwo{\ensuremath{\mathsf{\id_{c2}}}}
\newcommand\idhwOne{\ensuremath{\mathsf{\id_{hw1}}}}
\newcommand\idhwTwo{\ensuremath{\mathsf{\id_{hw2}}}}

\newcommand\securityPattern{\ensuremath{\mathsf{securityPattern}}}
\newcommand\safetyPattern{\ensuremath{\mathsf{safetyPattern}}}
\newcommand\safetyPatternID{\ensuremath{\mathsf{\id}}}
\newcommand\safetyPatternNAME{\ensuremath{\mathsf{name}}}
\newcommand\safetyPatternCP{\ensuremath{\mathsf{cp}}}
\newcommand\safetyPatternINP{\ensuremath{\mathsf{inp}}}
\newcommand\safetyPatternINT{\ensuremath{\mathsf{int}}}
\newcommand\safetyPatternOUT{\ensuremath{\mathsf{out}}}
\newcommand\safetyIntent{\ensuremath{\mathsf{safetyIntent}}}
\newcommand\securityIntent{\ensuremath{\mathsf{securityIntent}}}

\newcommand\idpt{\ensuremath{\mathsf{\id_{pt}}}}
\newcommand\idt{\ensuremath{\mathsf{\id_{th}}}}
\newcommand\idsg{\ensuremath{\mathsf{\id_{sg}}}}
\newcommand\mcs{\ensuremath{\mathsf{mcs}}}
\newcommand\idmcs{\ensuremath{\mathsf{\id_{msc}}}}
\newcommand\lmcsToHz{\ensuremath{\mathsf{lmcs2hz}}}
\newcommand\idlmcs{\ensuremath{\mathsf{\id_{lmcs}}}}

\newcommand\asil{\ensuremath{\mathsf{asil}}}
\newcommand\asila{\ensuremath{\mathsf{a}}}
\newcommand\asilb{\ensuremath{\mathsf{b}}}
\newcommand\asilc{\ensuremath{\mathsf{c}}}
\newcommand\asild{\ensuremath{\mathsf{d}}}

\newcommand\publicdsl{\ensuremath{\mathsf{public}}}
\newcommand\potThreat{\ensuremath{\mathsf{pThreat}}}
\newcommand\threatdsl{\ensuremath{\mathsf{threat}}}
\newcommand\reachI{\ensuremath{\mathsf{reachI}}}
\newcommand\pathI{\ensuremath{\mathsf{P}}}
\newcommand\bdCh{\ensuremath{\mathsf{bdCh}}}
\newcommand\ttpdsl{\ensuremath{\mathsf{pt_{tp}}}}
\newcommand\trtype{\ensuremath{\mathsf{th_{tp}}}}
\newcommand\severe{\ensuremath{\mathsf{sev}}}
\newcommand\moderate{\ensuremath{\mathsf{mod}}}
\newcommand\negligible{\ensuremath{\mathsf{neg}}}
\newcommand\authenticity{\ensuremath{\mathsf{auth}}}
\newcommand\availability{\ensuremath{\mathsf{avl}}}
\newcommand\integrity{\ensuremath{\mathsf{int}}}
\newcommand\confidentiality{\ensuremath{\mathsf{con}}}

\newcommand\hls{\texttt{HLS}\xspace}
\newcommand\cam{\texttt{C\scriptsize{AM}}\xspace}
\newcommand\cell{\texttt{C\scriptsize{ELL}}\xspace}
\newcommand\bt{\texttt{B\scriptsize{T}}\xspace}
\newcommand\navig{\texttt{N\scriptsize{AVIG}}\xspace}
\newcommand\gw{\texttt{G\scriptsize{W}}\xspace}
\newcommand\obd{\texttt{OBDC\scriptsize{onn}}\xspace}
\newcommand\bdctl{\texttt{B\scriptsize{D}\normalsize{C}\scriptsize{TL}}\xspace}
\newcommand\hlswt{\texttt{H\scriptsize{L}\normalsize{S}\scriptsize{WT}}\xspace}
\newcommand\pwrswtact{\texttt{P\scriptsize{WR}\normalsize{S}\scriptsize{WT}}\xspace}
\newcommand\canbus{\texttt{CAN Bus}\xspace}
\newcommand\can{\texttt{CAN}\xspace}
\newcommand\canI{\texttt{CAN Bus 1}\xspace}
\newcommand\canII{\texttt{CAN Bus 2}\xspace}
\newcommand\canIII{\texttt{CAN Bus 3}\xspace}
\title{Automating Safety and Security Co-Design through Semantically-Rich Architecture Patterns}


\author{Yuri Gil Dantas}
\affiliation{%
  \institution{fortiss GmbH}
  \city{Munich}
  \country{Germany}}
\email{dantas@fortiss.org}

\author{Vivek Nigam}
\affiliation{%
  \institution{Federal University of Para\'iba \& Huawei Technologies D\"{u}sseldorf GmbH}
  \city{Jo\~ao Pessoa \& D\"{u}sseldorf}
  \country{Brazil \& Germany}
}


\renewcommand{\shortauthors}{Dantas and Nigam}

\begin{abstract}
During the design of safety-critical systems, safety and security engineers make use of architecture patterns, such as Watchdog and Firewall, to address identified failures and threats.
Often, however, the deployment of safety patterns has consequences on security, \eg, the deployment of a safety pattern may lead to new threats.
The other way around may also be possible, \ie, the deployment of a security pattern may lead to new failures. 
Safety and security co-design is, therefore, required to understand such consequences and trade-offs, in order to reach appropriate system designs.
Currently, pattern descriptions, including their consequences, are described using natural language.
Therefore, their deployment in system design is carried out manually, thus time-consuming and prone to human-error, especially  given the high system complexity.
We propose the use of semantically-rich architecture patterns to enable automated support for safety and security co-design by using Knowledge Representation and Reasoning (KRR) methods.
Based on our domain-specific language, we specify reasoning principles as logic specifications written as answer-set programs.
KRR engines enable the automation of safety and security co-engineering activities, including the automated recommendation of which architecture patterns can address failures or threats and consequences of deploying such patterns.
We demonstrate our approach on an example taken from the ISO 21434 standard.
\end{abstract}

\ccsdesc[100]{Computer architectures~Safety\&Security}
\ccsdesc[100]{Logic~Knowledge Representation and Reasoning}


\maketitle

\section{Introduction}\label{sec:introduction}

Safety-critical systems are systems whose failure may result in severe consequences to human life, including death~\cite{Knight02}.
Examples of safety-critical systems are autonomous vehicles, and aircraft flight control.
The challenge for engineers is to ensure that such systems are safe at all times by, \eg, providing protective measures to reduce the risk of failures to an acceptable level.

This challenge increases substantially with the interconnectivity of safety-critical systems.
For example, vehicle platoons share information about their speed or position with other vehicles through wireless communication to enable vehicles to quickly react to sudden speed reductions.
The system interconnectivity brings security to the development life cycle of safety-critical systems, as an intruder might cause catastrophic events by remotely disabling safety functions.
Intruders can attack such communication channels to infiltrate vehicles and, \eg, disable safety functions thus reducing passenger safety~\cite{attack.jeep} or even causing accidents~\cite{dantas20vnc}.

These types of attacks have served as motivation for the new ISO 21434 standard for Automotive Cyber-Security~\cite{iso21434}. 
The standard also advocates a closer alignment between safety and security in order to ensure vehicle safety. 
That is, it advocates interactions between safety and security to coordinate the exchange of relevant information such as threat scenarios and hazard information or where a security requirement might conflict with a safety requirement.
These interactions are part of \emph{safety and security system co-design}, where trade-offs between safety and security are well understood, and optimal system designs are reached.

During system design, safety and security engineers deploy architecture patterns, \ie, patterns that are known to provide some type of guarantee for safety, \eg, fault tolerance, and security, \eg, separation.
Examples of safety patterns include Watchdog and Monitor Actuator, and examples of security patterns include Firewall and Security Monitor.

Currently, however, patterns are documented in a rather informal fashion~\cite{Armoush2010,preschern13plop,963314,SljivoUPG20} using natural language.
Therefore, it is the job of the safety and security engineers to correctly understand the textual description of patterns and propose manually the use of a particular pattern at a particular location of the system architecture.
As system complexity grows, this task becomes more complicated. This is because often patterns used for one aspect, may have consequences to other aspects~\cite{cambacedes13ress}.
Moreover, these consequences are context dependent. 
It may be that placing the same type of pattern in one location of the system architecture may have serious consequences, while placing the same pattern in another locations of the system architecture may not have these consequences.
For example, placing a firewall at a communication channel with safety-relevant information may unintentionally block safety-critical flows, while placing a firewall at a communication channel without safety-relevant information does not have safety consequences.
A second complicating factor is the correct understanding under which conditions a pattern can be used for attaining a safety goal or mitigating a security threat. 
For example, placing a safety pattern that adds heterogeneous redundancy, \ie, different implementations for the primary and secondary channels, instead of homogeneous redundancy, \ie, same implementation for both channels, 
propagates assumptions on the independence or not between primary and secondary channels. 
These assumptions need to be carefully understood during the development of the system.

Instead of describing patterns informally using natural language, a better approach
is to provide more precise semantics by using Domain-Specific Languages (DSL).\footnote{The semantics provided by DSL is not formal semantics, \ie, set of traces, but rather lightweight semantics, \ie, defining a vocabulary for which the meaning is uniformly understood in the corresponding domain. For example, it is clear in the automotive domain what ECU and CAN buses are.}
\emph{Semantically-rich architecture pattern} descriptions enable increased automated support during system design. 
In our previous work~\cite{dantas20iclp}, we demonstrate how system safety design can be automated by using semantically-rich safety patterns encoded as \emph{knowledge bases}~\cite{baral.book}, \eg, checking whether a safety pattern placed in the system architecture can be correctly used to attain a safety goal.

This paper's main goal is to enable safety and security co-design automation by using semantically-rich architecture patterns. 
Our main contributions are as follows:
\begin{itemize}
	\item \textbf{Domain-Specific Language (DSL):} We considerably extend our previous work~\cite{dantas20iclp} with a DSL for security, and for safety and security co-design. Moreover, we improve our DSL for safety to, \eg, more precisely specify the intent of safety patterns.
	\item \textbf{Semantically-Rich Safety and Security Patterns:} We propose safety and security patterns template that contains semantic information provided by the proposed DSL. Due to space restrictions, we describe in the paper only four such patterns, two for safety and two for security. Notice that our machinery (described below) currently supports ten patterns~\cite{safpat}. The supported patterns are Acceptance Voting, Homogeneous Duplex, Heterogeneous Duplex, Monitor Actuator, Simplex Architecture, Triple Modular Redundancy, Watchdog, Firewall and Security Monitor.  
	\item \textbf{Reasoning Principles:} We extend the safety reasoning principles proposed in our previous work~\cite{dantas20iclp} increasing their precision by, \eg, considering when a safety pattern fails operational. We specify security reasoning principles based on formal intruder models, \eg, path reachability used to determine whether an attack can pose a threat to some sub-component from outside the system, and when a security pattern can be used to mitigate some types of threats. We specify safety and security co-design reasoning rules, \eg, conditions for when a security pattern may cause safety failures and when safety patterns can be targeted by intruders so to reduce the system safety.  
	\item \textbf{Automated Reasoning:} We automate our machinery by using the off-the-shelf solver DLV~\cite{dlv}. 
	It enables the automated safety, security co-design with patterns. 
	We demonstrate this by using an example taken from the ISO 21434~\cite{iso21434}. We refer to the whole machinery proposed in this paper by \machinery. \machinery is capable of automating several activities that are currently carried out manually. \machinery is publicly available in~\cite{safpat}.

\end{itemize}

The remainder of this paper is organized as follows:
Section~\ref{sec:preliminaries} reviews basic safety and security concepts, the V-model used in vehicle system development, and templates used for describing patterns. 
Section~\ref{sec:headlamp} describes the running example taken from ISO 21434. It also illustrate the main artifacts constructed during the execution of the V-model.
Section~\ref{sec:framework} describes our DSL for safety, security, and safety and security co-analysis.
Sections~\ref{sec:safety} and \ref{sec:security} demonstrate with some examples how to semantically enrich patterns 
using the proposed DSL. 
Section~\ref{sec:automation} demonstrates how safety and security co-design can be supported by semantically-rich architecture patterns by automation through DLV.
Finally, we conclude by pointing out to related and future work in Sections~\ref{sec:related-work} and \ref{sec:conclusion}.
\section{Safety and Security Concepts, V-Model, and Patterns}\label{sec:preliminaries}

\begin{figure}
  \includegraphics[width=0.8\textwidth]{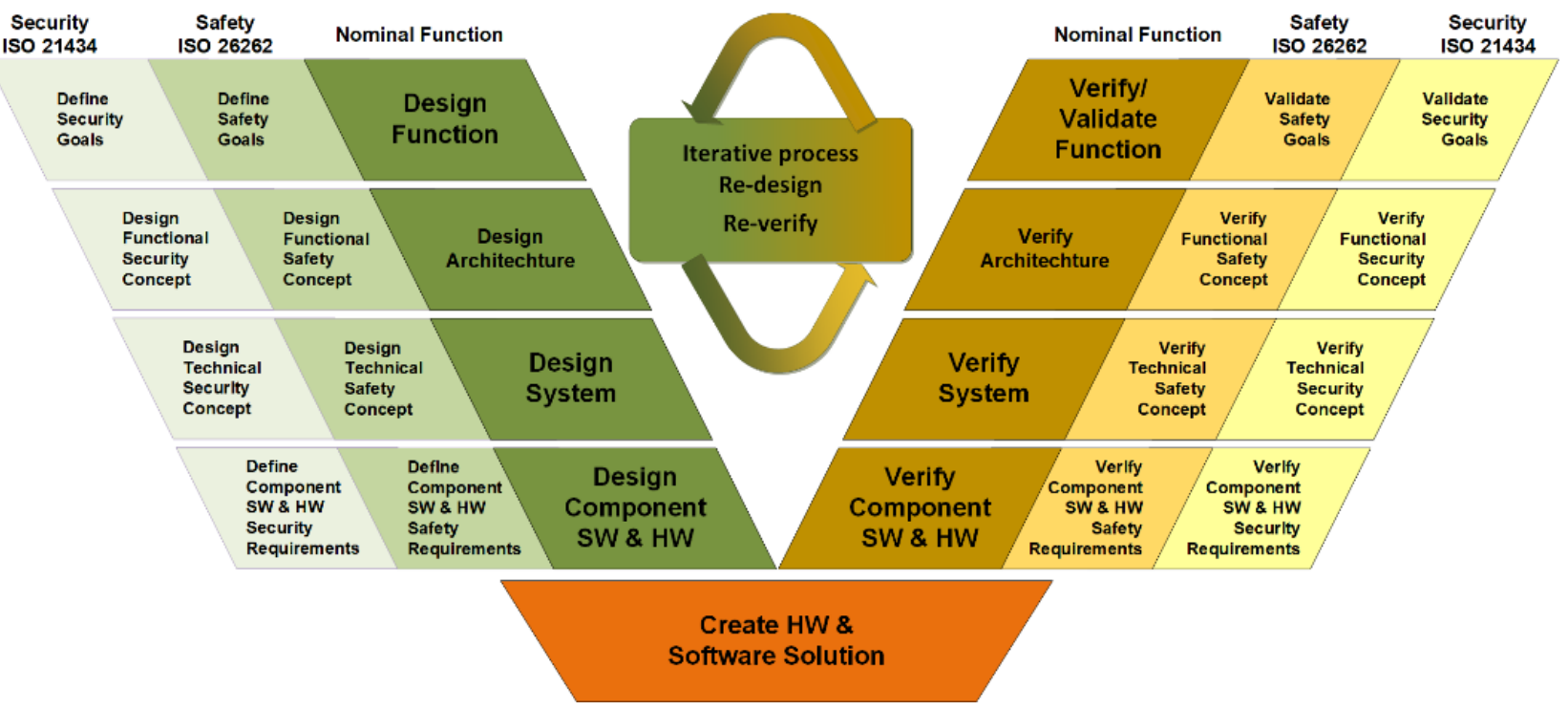}
  \caption{Integrated V-Model for System, Safety~\cite{skoglund18assure}.}
  \label{fig:v-model}
\end{figure}

The process, methods and artifacts that shall be produced during the development of vehicle embedded systems are detailed in the standards ISO 26262~\cite{iso26262} for safety and ISO 21434~\cite{iso21434,dantas-whitepaper2020} for security. 
The overall process follows the so-called V-model, shown in Figure~\ref{fig:v-model}.

Before we enter into the details of the process, we briefly review some basic concepts in safety and security to set the terminology used in the remainder of the paper. For both safety and security, an \emph{item} is the system or combination of systems to implement a function at the vehicle level.

\paragraph{Basic Safety Concepts:} The definition of the following safety concepts are taken in their great majority from~\cite{AvizienisLRL04}.
A \emph{hazard} is a situation that can cause harm to users or businesses. 
A \emph{failure} is an event that when occurs results in a deviation of the expected behavior of a function. 
An \emph{error} is a deviation of the expected system behavior. 
A \emph{fault} is the hypothesized cause of an error. 
A \emph{Minimal Cut Set} (MCS) is a set of failures that when occurring at the same time (or in sequence) may lead to a (top-level) failure. 
This top-level failure is often associated with a hazard using techniques such as Hazard and Operability study (HAZOP). 
Normally failures are associated with a set of predefined Guidewords that characterize intuitively the semantics of such failures.
Examples of Guidewords are \emph{loss} and \emph{erroneous} that denote, respectively, a failure due to the loss of a function, \ie, a function not operating at all, and a failure due to an erroneous function behavior, \eg, a function not computing correctly some output value.
The MCSes of a top-level failure are typically computed from a Fault Tree Analysis (FTA), which is a deductive safety analysis method that decomposes failures using an and/or tree of sub-failures. 
\emph{Fault tolerance} are means to avoid service failures in the presence of faults. 
A \emph{safety pattern} (described in further details in Section~\ref{subsec:patterns}) is a system architecture solution that is known to provide some level of fault-tolerance.   

\paragraph{Basic Security Concepts:} The definition of the following security concepts are taken in their great majority from the ISO 21434~\cite{iso21434}. 
An \emph{asset} is an object for which the compromise of its cybersecurity properties can lead to damage for an item's stakeholder. 
A \emph{damage scenario} is an adverse consequence involving a vehicle or vehicle function (\eg, an asset) and affecting a road user.
An \emph{attack} is an deliberate action or interaction with the item or component or its environment that has potential to result in an adverse consequence. 
A \emph{threat scenario} is a statement of potential negative actions that lead to a damage scenario. 
An \emph{attack path} is a sequence\footnote{The ISO 21434 define a path as a set and not as sequence. We will use here as a sequence.} of actions that could lead to the realization of a threat scenario. 
A \emph{cybersecurity property} is an attribute of an asset including confidentiality, integrity and availability.
\emph{A cybersecurity risk} is the effect of uncertainty on road vehicle security expressed in terms of \emph{attack feasibility} and \emph{impact}.
A \emph{security pattern} is a system architecture solution that is known to control security risks by mitigating threat scenarios.

\smallskip

At the top left of the V-model shown in Figure~\ref{fig:v-model}, one defines the item that will be subject of safety and security analysis by providing details such as the item's preliminary architecture and operational conditions. 
Then safety and security analysis are performed
in order to define safety and security goals (top left boxes in the left-branch of the V-model depicted in Figure~\ref{fig:v-model}). 
Safety analyses, such as Hazard Analysis and Risk Assessment (HARA), FTA, HAZOP, identify losses, hazards, failures leading to hazards, and faults that may trigger such failures.  Security analysis, such as Threat Agent Risk Assessment (TARA), STRIDE\footnote{The threats considered by STRIDE are Spoofing, Tampering, Repudiation, Information disclosure, Denial of service, and Elevation of privilege.}, Attack Trees~\cite{shostack14book} and path analysis~\cite{iso21434}, identify key assets and their corresponding threats.
Security risks are determined by assessing the attack feasibility and impact of each identified attack path in the system leading to a threat scenario.
Typically, an attack path feasibility is evaluated by using different factors such as the time needed to carry out an attack and the knowledge/tooling needed by the intruder. 
Moreover, different categories for impact may be considered, such as financial, privacy, and safety. Safety related impacts have the highest impact.   

Based on the risk evaluation of such hazards and threats, functional safety and security concepts are formulated for the item by establishing safety goals/key risks on the system architecture~\cite{iso26262} (second boxes in the left-branch of the V-model depicted in Figure~\ref{fig:v-model}).

The safety functional concept establishes the level of criticality required by elements in the system. 
For safety, vehicle functions are assigned values between ASIL QM, A, B, C, D, where QM has no safety criticality, while A,B,C,D have ascending criticality requirements.
Safety goals may also assign the level of tolerance to faults for functions as defined below:
\begin{itemize}
  \item \textbf{Fail-Silent} is a more strict type of fault tolerance when compared to fail active (function fails without any measure) as the failure of fail silent function shall necessarily lead to the loss of the functionality (\eg, shutdown the function). Thus, it shall not be possible that the faults of a fail silent function, \eg, incorrect computations, are propagated within the system. 
  
  \item \textbf{Fail-Safe} adds the requirement to fail-silent in that if a function fails, then it shall necessarily switch to a safe-state. 
  Moreover, as we will describe below with the architecture patterns, it is possible to effectively detect when a function is not longer functioning and switch to a safe mode and trigger appropriate measures, \eg, inform the driver.

  \item \textbf{Fail-Operational} is the strongest type of fault tolerance as it requires that a function operates with the same level of safety even after facing a specific number of faults.
  For example, a lane keeping function typically shall operate complying to the requirements of the highest level of criticality, \ie, ASIL D, after at least one fault occurred.   
\end{itemize}

The security goals establish the properties, \eg, confidentiality, integrity and availability, that need to be satisfied by the identified assets in the system architecture.

\subsection{Safety and Security Architecture Patterns}
\label{subsec:patterns}
Once the functional safety and security concepts are completed, 
the technical safety and security concepts are developed (third boxes in the left-branch of the V-model depicted in Figure~\ref{fig:v-model}).
This is done by further establishing safety and security requirements on the item system architecture, so to comply, for example, with the level of safety criticality established and security properties required. 
Typically, safety and security engineers make use architecture solutions, called architecture patterns.

\emph{Architecture patterns} are abstract solutions to recurrent system problems such as safety and security.\footnote{Other measures like testing and established coding practices may be used in addition or instead of architecture patterns.}
For example, safety engineers make use of a Triple Modular Redundancy to address failures (both erroneous and losses) thus avoiding hazards.
Similarly, security engineers use firewalls to isolate the system architecture thus reducing security risks. 
These architecture patterns are described in an abstract form and they are independent of implementation details. 
Their description make them easier to understand and can be seen as guidelines for the design of the system architecture. 
It is not the part of architecture patterns to define exactly how patterns components shall be implemented, although requirements might be provided. 
The actual implementation of pattern components such as monitors shall be tailored to specific functions. For example, monitors are often implemented using plausibility checks (tailored to specific functions).
A collection of known patterns is available in the ISO 26262~\cite{iso26262} as well as in literature~\cite{Armoush2010,preschern13plop}.

\begin{table}[h]
\centering
\begin{tabular}[t]{p{4cm}p{8.5cm}}
\toprule Field
&Description \\
\midrule
Pattern name& Name of this pattern.\\
\hline
Structure& Block diagram of this pattern.\\
\hline
Intent& Textual description of the purpose of this pattern.\\
\hline
Problem addressed& Textual description of the problem the pattern addresses.\\
\hline
Assumptions (requirements) & Assumptions necessary for using this pattern.\\
\hline
Consequences& Textual description of the consequences to other concerns, \eg, security, performance, reliability. \\
\bottomrule
\end{tabular}
\caption{Architecture Pattern Description Template}
\label{table:pat-template}
\end{table}%

Since patterns are commonly used, and many times using different names, 
pattern templates have been proposed~\cite{Armoush2010,SljivoUPG20,preschern13plop} as a means to uniformly describe 
a pattern. 
An example of a template is depicted in Table~\ref{table:pat-template} which is similar to pattern templates appearing in the literature~\cite{Armoush2010,SljivoUPG20}.
The description of the pattern contains a name for which a pattern is known for; its structure, typically shown as a block diagram; its intent, \ie, the purpose for which this pattern is normally used, \eg, to enable fail-operational level of fault tolerance; problem addressed, \eg, to control erroneous functions or loss of functions; assumptions (a.k.a. requirements) required to use this pattern, \eg, different types of implementations for the primary and redundant channels; consequences of using this pattern to other concerns, such as security, performance, reliability, costs.

\section{Running Example}\label{sec:headlamp}

This section describes an example from the automotive domain, namely headlamp system, taken from the ISO 21434 standard~\cite{iso21434}.
We use this example to illustrate the concepts and process reviewed in Section~\ref{sec:preliminaries}, as well as to illustrate the machinery introduced in the next sections.
Following the process described in Section~\ref{sec:preliminaries}, we start by providing a description of both the item, \ie, headlamp system, as well as the results of a safety analysis of the headlamp system.

\begin{figure}[h]
 \centering
 \includegraphics[width=0.8\textwidth]{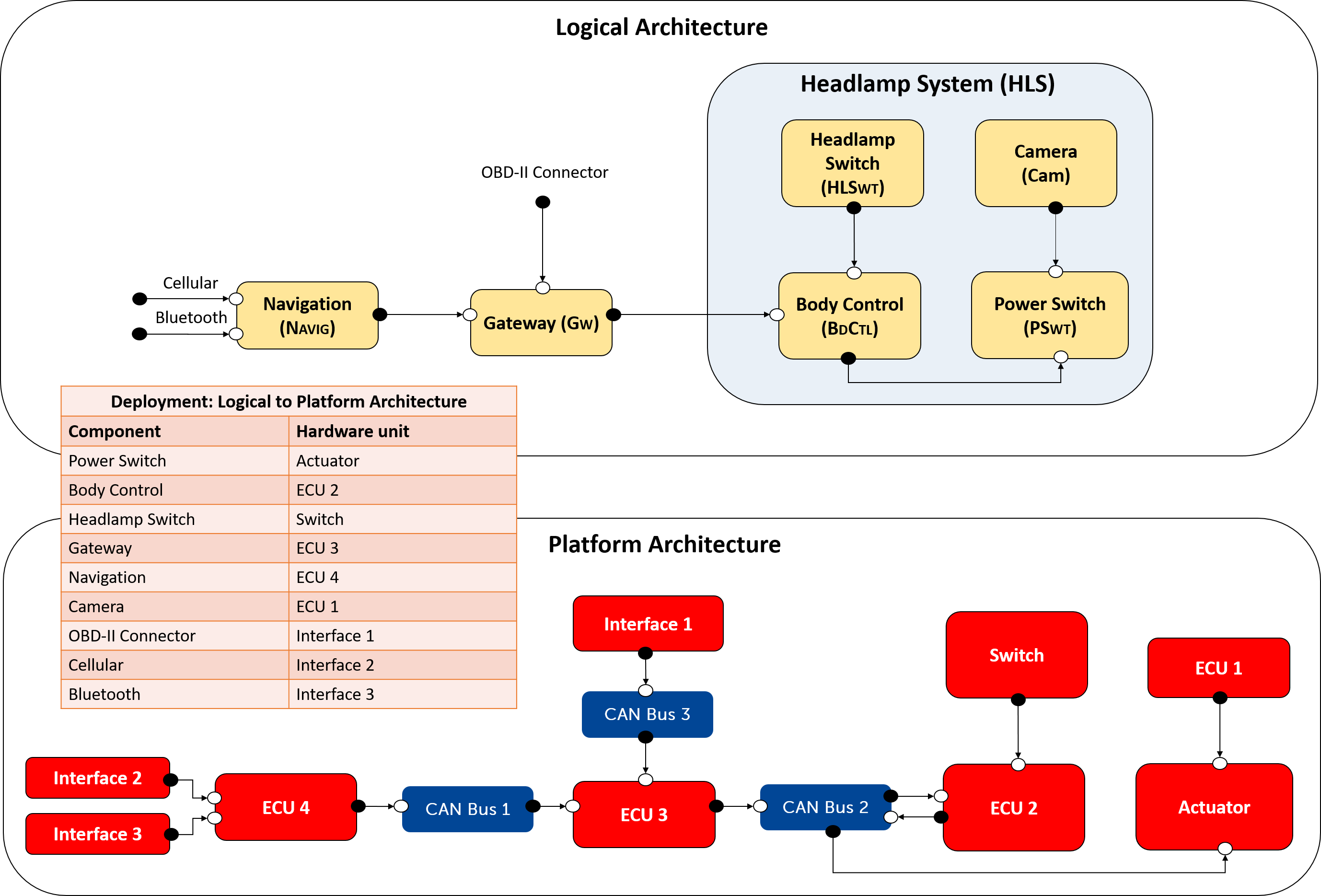}
 \caption{Architecture of the headlamp system}
 \label{fig:hls-architecture}
 \Description{}
\end{figure}

\paragraph{Functionalities and System Architecture} 
A headlamp system is responsible for switching on/off the headlamp of a vehicle.
The headlamp system has two specific features, namely high-beam light and low-beam light. 
The driver can turn on or off the headlamp and switch between high-beam and low-beam from the steering wheel.
Since high-beam may affect the visibility of drivers incoming from the opposite direction, 
it is a safety recommendation that a vehicle's headlamp is switched to low-beam whenever 
an oncoming vehicle is approaching from the opposite direction.
A proposed solution is to use a sensor, \eg, a camera, to detect approaching vehicles.
If the headlamp is in high-beam mode, the headlamp system switches the headlamp automatically to low-beam mode when an oncoming vehicle is detected. 
It returns automatically the headlamp to high-beam mode if the oncoming vehicle is no longer detected~\cite{iso21434}. 

Figure~\ref{fig:hls-architecture} depicts both the logical and the platform architecture of the headlamp system, as well as the deployment table of the logical architecture to the platform architecture.
The boxes in the logical architecture represent components, \eg, \pwrswtact, \bdctl, and the boxes in the platform architecture represent hardware units, \eg, Interface 1, \canII.
The black and white circles connected to components/hardware units are, respectively, output and input ports.
The arrows connected to ports represent unidirectional channels between components/hardware units.

A Camera (\cam) detects oncoming vehicles and sends signals to the Power Switch (\pwrswtact).
The Body Control (\bdctl) sends signals to \pwrswtact.
These signals are requests from the driver (coming from the Headlamp Switch -- \hlswt) to turn the headlamp on or off.
Note that the channel from \bdctl to \pwrswtact is deployed to a CAN bus (\canII) in the platform architecture.
The left-hand side of the logical architecture depicts external components that may access the headlamp system.
There is a Gateway (\gw) that control access from other components located outside the headlamp system, e.g., Navigation (\navig).
\gw receives signals from an OBD-II Connector (\obd).
\navig has two interfaces, namely Cellular (\cell) and Bluetooth (\bt).
Both \cell and \bt interfaces and \obd may access \bdctl to, \eg, carry out software updates.
Note that in the platform architecture the channel from \navig to \gw is deployed to \canI, the channel from \gw to \bdctl is deployed to \canII, and the channel from \obd to \gw is deployed to \canIII.

\paragraph{Safety Analysis Results}
We now focus on the safety of the headlamp system.
While not claiming to be comprehensive, we describe potential hazards, faults, and failures that can be identified from a safety analysis.
We also describe safety goals that shall be met to address the identified hazards.


Table~\ref{table:hls-hazards} describes the identified hazards for the headlamp system. 
The ASIL level of a hazard is assigned based on three parameters: Severity, Exposure and Controllability~\cite{iso26262}.
\emph{Severity} denotes the consequences to the life of the user of the system in the presence of a failure that leads to the hazard.  
\emph{Exposure} denotes the possibility of the system being in a hazardous situation that can cause harm.
\emph{Controllability} denotes the extent to which the user of the system can control the system in the presence of a failure that leads to the hazard.
We assign ASIL C to \textbf{HZ1}: We consider the severity as life threatening injuries (S3), the exposure as medium probability of happening (E3), and the controllability as difficult to control (C3).
We assign ASIL A to \textbf{HZ2}: We consider the severity as light and moderate injuries (S1), the exposure as high probability of happening (E4), and the controllability as normally controllable (C2).

\begin{table}[t]
  \begin{tabular}{p{1cm}p{10.5cm}p{1cm}}
    \toprule
    \textbf{Hazard} & \textbf{Description} & \textbf{ASIL}\\
    \midrule
    HZ1 & Headlamp turn off unintentionally during night driving. & C \\
    \midrule
    HZ2 & Unintended low beam of headlamp when no oncoming vehicle is detected. & A \\
    \bottomrule
  \end{tabular}
  \caption{Identified hazard for the headlamp system}
  \label{table:hls-hazards}
\vspace*{-0.8cm}
\end{table}

We list below the identified faults and failures that may lead to the presence of hazards \textbf{HZ1} and \textbf{HZ2}.
Specifically, we consider failures of type erroneous and loss.

\begin{itemize}
	\item \textbf{FT1:} The Body Control is faulty thus leading to not turning the headlamp on upon the driver's request. This may happen if fault \textbf{FT1} triggers a failure \textbf{FL1} of type erroneous. The failure \textbf{FL1} may lead to hazard \textbf{HZ1}.
	\item \textbf{FT2:} The logical channel between the Body Control and the Power Switch is faulty, leading to not turning the headlamp on upon the driver's request. This may happen if fault \textbf{FT2} triggers a failure \textbf{FL2} of type loss. The failure \textbf{FL2} may lead to hazard \textbf{HZ1}.
	\item \textbf{FT3:} The Camera is faulty, not providing the expected information to the Power Switch to enable the high-beam light. This may happen if fault \textbf{FT3} triggers a failure of type erroneous. The failure \textbf{FL3} may lead to hazard \textbf{HZ2}.
\end{itemize}
Since all of these failures can lead independently to the hazard, the minimal cut sets are \{\textbf{FL1}\}, \{\textbf{FL2}\}, and \{\textbf{FL3}\}.

Table~\ref{table:hls-safety-goals} describes a safety goal to address \textbf{HZ1}.
Notice that safety goals are often expressed as a negation of a hazard. 
Here, we consider a more specific safety goal to enable the reasoning of safety patterns. 
The safety goal \textbf{SG1} aims at avoiding potential harm always after the 1st failure of type erroneous. 
The system shall transition to a safe state always after the 2nd failure.
If achieved, this safety goal improves both the safety and availability of the headlamp system.
\textbf{SG1} can be achieved by the implementation of safety patterns, implementing, \eg, fault tolerance tactics.
\textbf{SG2} describes a safety goal that allows the system to fail silent due the low criticality of hazard \textbf{H2}.
Note that we neglect fault \textbf{FT2}, as we consider that this fault is unlikely to happen.

\begin{table}[t]
  \begin{tabular}{p{1.8cm}p{8.5cm}p{1cm}p{1cm}}
    \toprule
    \textbf{Safety Goal} & \textbf{Description} & \textbf{ASIL} & \textbf{Hazard} \\
    \midrule
    SG1 & The system shall fail operational always after the 1st erroneous failure of the Body Control. The system shall transition to a safe state always after the 2nd failure of type erroneous. & C & HZ1\\
    \midrule
    SG2 & The system shall fail silent always after most 1st erroneous failures on the Camera. & A & HZ2\\
    \bottomrule
  \end{tabular}
  \caption{Safety goals to prevent the presence of identified hazards}
  \label{table:hls-safety-goals}
  \vspace*{-0.8cm}
\end{table}

\paragraph{Security Analysis Results}
For demonstrating system safety, one shall argue that the defined safety goals are met.
Since the headlamp system is safety critical, it is considered an asset. 

The next step is to determine the threats to the headlamp system. 
Failures shall immediately be considered as threats as recommended by the ISO 26262~\cite{iso26262} and the ISO 21434~\cite{iso21434}.\footnote{Notice, however, that security shall also consider threats that are not directly safety-related, such as threat posed to privacy.  Since our focus is on safety, we do not focus on these types of threats in this paper.}
For example, one shall evaluate the risk of attacking the Body Control to cause it to fail. 
Notice that the intruder can also cause the CAN to fail. 
So, although from a safety perspective a CAN failure is very rare, from a security perspective
a CAN denial of service attack can easily be carried out provided the intruder can access that CAN.
Section~\ref{ssec:automation-security-patterns} demonstrates how the association of failures and threats can be derived by reasoning rules.

Once the threats are identified, one carries out a risk analysis.
This is done by enumerating the attack paths leading to threats to the headlamp.
To compute the attack paths, one identifies which are the platform architecture elements from which the intruder can 
access the system. We classify such elements as \emph{public}.
Consider the platform architecture of the headlamp system. 
We consider the following hardware units as public elements: Interface 1 (OBD-II Connector), Interface 2 (Cellular) and Interface 3 (Bluetooth).
That is, an attack may access each of these hardware units to carry out attacks against the headlamp system.
The exact attack path depends on the threat model considered. 
In Section~\ref{ssec:automation-safety-security-patterns}, we consider a threat model based on the Dolev-Yao intruder~\cite{DY} used in protocol security, where the intruder
attempts logical attacks to the identified assets. 
Based on this threat model, the attack paths are enumerated. 

Finally, security requirements with security countermeasures, \eg, security patterns, 
are proposed to mitigate the identified high risk threats.

\section{Knowledge Bases for Safety and Security System Architecture}\label{sec:framework}

Our goal is to provide mechanisms to support the automated hardening of system architectures using safety and security patterns.
To this end, we propose the use of Knowledge Representation and Reasoning (KRR) methods~\cite{baral.book} revisited in Section~\ref{ssec:kb-asp}. 
KRR enables the specification of sophisticated reasoning principles that can be automated by reasoner tools such as DLV. 

We have recently proposed the use of KRR for hardening system architectures using safety patterns~\cite{dantas20iclp}.
The main outcome of this work was \machineryOld.
\machineryOld consists of a Domain-Specific Language (DSL) for embedded systems, and safety reasoning principles for some selected safety patterns specified as disjunctive logic programs. Finally, we demonstrated how \machineryOld can recommend safety patterns in an automated fashion. 

This paper substantially extends the previously developed \machineryOld.
More concretely, we extend \machineryOld \cite{dantas20iclp} in the following ways:
\begin{itemize}
	\item We extend our DSL to specify minimal cut sets, faults, failures, and safety goals, thus following more closely the process described in Section~\ref{sec:preliminaries}.
	\item We extend our DSL to more precisely specify safety patterns, including when a safety pattern enables a system to fail operational, fail safe, and fail-silent, and which safety patterns are suitable for addressing life critical (ASIL C and D) and low critical (ASIL A and B) hazards. 
	\item The main extension of \machineryOld is the introduction of security aspects to enable the automated recommendation of security patterns. Since \machineryOld now includes security aspects (in addition to safety aspects), we changed its name from \machineryOld to \machinery.
	\item Another important extension is the introduction of security consequences when applying safety patterns, and safety consequences when applying security patterns.
	\item We specify reasoning principles to specify assumptions required to use patterns.
	\item We specify constraints to limit the number of architecture solutions with patterns recommended by \machinery.
\end{itemize}	

The focus of this section is on the DSL of \machinery. 
Sections~\ref{sec:safety} and \ref{sec:security} describe, respectively, our specification of safety and security patterns by example.
Section~\ref{sec:automation} describes our reasoning principles that are automated by DLV.
Next, we provide a brief overview on Knowledge Representation and Reasoning to help readers to grasp the developed \machinery.

\subsection{Knowledge Bases and Answer-Set Programming}\label{ssec:kb-asp}
Knowledge Representation and Reasoning (KRR)~\cite{baral.book} is a mature field of Artificial Intelligence based on logic-based methods to represent and reason about knowledge bases. 
A \emph{knowledge base} is a declarative representation of the world/system. 
The declarative nature of knowledge bases enables the  programming of reasoning rules by using existing logic programming engines such as DLV~\cite{dlv}.

A disjunctive logic program $M$ is a set of rules of the form $a_1 \lor \cdots \lor a_m \leftarrow \ell_1, \ldots, \ell_n$ or 
 where $\ell, \ell_1, \ldots, \ell_n$ are literals, that is atomic formulas, $a$, or negated atomic formulas ${not}~a$.
The interpretation of the default negation $not$ assumes a \emph{closed-world} assumption. That is, we assume to be true only the facts that are explicitly supported by a rule.

The semantics of a disjunctive logic program $M$ is based on the stable model semantics~\cite{gelfond90iclp}. We illustrate the semantics of logic programs with an example.\footnote{We refer to~\cite{baral.book} for the precise formal semantics of logic programs.} 

Consider the program $P_1$ with the following two rules:
\begin{center}
\(
a \lor b \qquad c \leftarrow a
\)
\end{center}

\noindent
$P_1$ has two answer-sets  $\{a,c\}$ and $\{b\}$. Intuitively, each answer-set is a minimal model of the logic program, \ie, that makes each rule of the program to be true. Moreover, if a rule's  head is empty, \ie, $m = 0$ from the set of rules above, then it is a constraint. 
For example, if we add the clause $\leftarrow b$ to $P_1$, then the resulting program has only one answer-set $\{a,c\}$.

DLV~\cite{dlv} is an engine implementing disjunctive logic programs based on ASP semantics~\cite{gelfond90iclp}. 
In the remainder of this paper, we use the DLV notation writing \code{:-} for $\leftarrow$ and \code{v} for $\lor$, \eg, the program $P_1$ is written as \code{a v b} and \code{c :- a}. As with DLV, capital letters \code{X,Y,Z} are variables that during execution are instantiated by appropriate terms, and minuscule letters \code{a,b,c} are constants.  
Variables or constants surrounded by [] are lists.
The \_ (underscore) character specifies that the argument can be ignored in the current rule.

\subsection{A Domain-Specific Language for Embedded Systems}\label{sec:dsl}

\machinery consists of a Domain-Specific Language (DSL) for embedded systems.
This DSL enables the specification of architecture elements, safety and security elements, and architecture patterns.
These elements are specified by the user of \machinery, while the reasoning rules in Section~\ref{sec:automation} are under the hood.
We illustrate our DSL using the headlamp system described in Section~\ref{sec:headlamp}.


\subsubsection{Architecture Elements} Table~\ref{table:dsl-arch-elements} describes selected architecture elements specified in our DSL, including components, sub-components, channels, and information flows.

\begin{table}
  \begin{tabular}{p{2.7cm}p{11cm}}
    \toprule
    \textbf{Fact} & \textbf{Description}\\
    \midrule
    \cp(\id) & $\id$ is a component in the system.\\
    \midrule
    \subcp($\id_1$,$\id_2$) & $\id_1$ is a sub-component of component $\id_2$.\\
    \midrule
    \ch(\id,$\id_1$,$\id_2$) & $\id$ is a logical channel connecting an output of component $\id_1$ to an input of component $\id_2$.
    Notice that it denotes a unidirectional connection.\\
    \midrule
    $\info(\id,[\ch_1,\ldots,\ch_n])$ & $\id$ is an information flow following the channels in $[\ch_1,\ldots,\ch_n]$.\\
    \midrule
    \ecu(\id) & $\id$ is an Electronic Computing Unit (ECU) that can run components.\\
    \midrule
    \candsl(\id) & $\id$ is a Controller Area Network (CAN) to communicate between ECUs.\\
    \midrule
    \interface(\id) & $\id$ is a hardware interface that allows the connection of external peripheral to components.\\
    \midrule
    \Dep(\id,\idd) & Component $\id$ is deployed (\ie, executed) in ECU \idd\ or in interface \idd, or alternatively, the logical channel $\id$ is deployed in CAN \idd\ to establish the communication between $\id$'s components.\\
    \bottomrule
  \end{tabular}
  \caption{\machinery: \machinery: Language for (selected) architecture elements}
  \label{table:dsl-arch-elements}
  \vspace*{-0.8cm}
\end{table}

\begin{example}
Consider the architecture of the headlamp system described in Section~\ref{sec:headlamp}. 
A user may specify the logical architecture of the headlamp system as follows:

\begin{flushleft}
\small{
\code{cp(cam). cp(bdCtl). cp(ps). cp(hlSwt). cp(hls). cp(gw). cp(navig). cp(obdC). cp(cell).}\\
\code{cp(bt). subcp(cam,hls). subcp(bdCtl,hls). subcp(ps,hls). subcp(hlSwt,hls).}\\
\code{ch(cmpsa,cam,ps). ch(hsbd,hlSwt,bdCtl). ch(bcps,bdCtl,ps). ch(gwbc,gw,bdCtl).}\\ 
\code{ch(navgw,navig,gw). ch(obdgw,obdC,gw). ch(celnav,cell,navig). ch(btnav,bt,navig).}\\  
\code{if(if1,[cmps]). if(if2,[hsbd,bcps]). if(if3,[obdgw,gwbc,bcps]).}\\ 
\code{if(if4,[celnav,navgw,gwbc,bcps]). if(if5,[btnav,navgw,gwbc,bcps]).} }
\end{flushleft}

The facts \code{cp(hlSwt)}, \code{cp(bdCtl)}, \code{cp(ps)}, and \code{cp(hls)} denote, respectively, the Headlamp Switch, the Body Control, the Power Switch, and the Headlamp system components.
The fact \code{subcp(bdCtl,hls)} denotes that the Body Control is a sub-component of the Headlamp system.
The fact \code{ch(bcpsa,bdCtl,ps)} denotes the logical communication channel \code{bcpsa} between the Body Control and the Power Switch.
The information flow \code{if2} denotes data flows from channel \code{hsbd} to channel \code{bcps}.

A user may specify the hardware units of the the platform architecture as follows. 
We only consider the hardware units shown in Figure~\ref{fig:hls-architecture}, omitting, \eg, the communication medium between the Body Control and the Power Switch.

\begin{flushleft}
\small{
\code{ecu(ecu1). ecu(ecu2). ecu(ecu3). ecu(ecu4). can(can1). can(can2). wireless(wl).}\\
\code{interface(int2). actuator(act). interface(int1). interface(int3). can(can3).switch(swt). }}  
\end{flushleft}

The fact \code{ecu(ecu1)} denotes the ECU \code{ecu1}.
The facts \code{switch(swt)} and \code{actuator(act)} denotes the switch \code{swt} and the actuator \code{act}, respectively.
The fact \code{interface(int1)} denotes interface \code{int1}.
The fact \code{wireless(wl)} denotes a wireless communication apparatus identified as \code{wl}.

The deployment of the logical architecture to the platform architecture is specified as follows:

\begin{flushleft}
\small{
\code{dep(cam,ecu1) dep(bdCtl,ecu2). dep(gw,ecu3). dep(navig,ecu4). dep(navgw,can1).}\\
\code{dep(gwbc,can2). dep(bcps,can2). dep(obdgw,can3). dep(cell,int1). dep(bt,int2).}\\ 
\code{dep(obdC,int3). dep(celnav,wl). dep(btnav,wl).}}
\end{flushleft}

The fact \code{dep(bdCtl,ecu2)} denotes that the Body Control is deployed to ECU \code{ecu2}.
The fact \code{dep(gwbc,can2)} denotes that the channel from the Gateway to the Body Control is deployed to CAN bus \code{can2}.
The fact \code{dep(btnav,wl)} denotes that the data transmission of logical channel \code{btnav} is performed via wireless. 
\end{example}
\begin{table}
  \begin{tabular}{p{4.7cm}p{8.6cm}}
    \toprule
    \textbf{Fact} & \textbf{Description}\\
    \midrule
    \hz(\idhz,[\system,\hzsev,\hzexp,\hzctl]) & \idhz\ is a hazard for system \system\ of severity \hzsev, exposure \hzexp, and controllability \hzctl.\\
    \midrule    
    \ft(\idft,[\idc]) & \idft\ is a fault associated with component \idc.\\ 
    \midrule
    \fl(\idfl,[\tp]) & \idfl\ is a failure of type $\tp$, where $\tp \in \{\err,\loss\}$.\\
    \midrule
    \ftTofl(\idft,\idfl) & \idft\ is a fault that triggers failure \idfl.\\
    \midrule
	\mcs(\idmcs,[\idfl]) & \idmcs\ is a minimal cut set consisting of failure(s) \idfl. \\
	\midrule
	\lmcsToHz([\idmcs],\idhz) & \idmcs\ is a list of minimal cut sets that leads to hazard \idhz.\\
	\midrule	         		         
    \sg(\idsg,[\idhz,\fop,\fsilent,\fsafe]) & \idsg\ is a safety goal to address hazard \idhz. This safety goal requires the system to either fail operational (\fop), fail silent (\fsilent) or fail safe (\fsafe) in the presence of \idhz.\\

    \midrule
    \publicdsl(\idhw) & \idc\ is a HW unit that may be accessible by external users.\\
    \midrule
    \potThreat(\idpt,[\idc,\idhw,\ttpdsl,\sev]) & \idpt\ is a potential threat associated with component \idc\ and HW unit \idhw. \idpt\ is of type $\ttpdsl$, where $\ttpdsl \in \{\confidentiality,\integrity,\availability\}$, and of severity $\sev$, where $\sev \in \{\negligible,\major,\moderate,\severe\}$.\\
    \midrule
    \reachI(\idc,\idhw, \pathI) & component \idc\ and HW unit \idhw\ may be reached by an intruder through path $\pathI$ in the technical architecture.\\ 
    \midrule 
    \threatdsl([\idt,\pathI],[\idc,\idhw,\trtype,\trsev]) & \idt\ is a threat associated with component \idc\ and HW unit \idhw\ that may be reached through path \pathI. \idt\ is of type $\trtype$ and of severity $\trsev$ (both as for \potThreat).\\ 
    \bottomrule
  \end{tabular}
  \caption{\machinery: Language for safety and security elements}
  \label{table:dsl-safety-safety}
\end{table}

\subsubsection{Safety and Security Elements} Our DSL consists of safety and security elements described in Table~\ref{table:dsl-safety-safety}.
By \emph{safety elements}, we refer to safety goals, hazards, faults, failures, and minimal cut sets. 
By \emph{security elements}, we refer to potential threat and threat scenarios.~\footnote{Our DSL enables the specification of further security elements such as damage scenario and risk determination as shown in~\cite{dantas-whitepaper2020}. We omit these security elements for the sake of presentation given that our focus is on safety and security co-design using architecture patterns.}
A potential threat associated with a hardware unit \textbf{HWCP} becomes a threat if there is a path \textbf{P} from a public element to \textbf{HWCP}.
Table~\ref{table:dsl-safety-safety} also includes further elements (\eg, \code{ft2fl}, and \code{reachI}) needed to reason about safety and security as described in Section~\ref{sec:automation}.

Motivated by~\cite{963314}, we consider safety goals that require the system to either fail operational, fail silent, or fail safe.
This has a great impact on the precision of \machinery in recommending safety patterns, as, \eg, only a sub-set of safety patterns can ensure that the system fails operational.
Whether the system fails operational, silent, or safe is specified as parameters of the predicate \sg~(see Table~\ref{table:dsl-safety-safety}).  
Either of these parameters (\fop, \fsilent, \fsafe) may be assigned to the following constants:

\begin{center}
$\fop,\fsilent,\fsafe \in \{\allNFail, \mostNFail, \never\}$, where
\end{center}
\allNFail\ denotes always after X failures have been detected, where X is an integer. 
For example, the system shall fail operational always after the 1st failure has been detected. 
According to~\cite{mil-std-2165}\cite{963314}, safety patterns that implement plausibility checks (\eg, monitor-actuator pattern~\cite{Armoush2010}) can only detect about 95\% of the failures.
Hence, we consider the constant \mostNFail\ that denotes always after most X failures have been detected, where X is an integer.
For example, the system shall fail safe after most 1st failures.
This means that some of the failures will not be detected by the pattern and the system will not always fail safe.
This is an important distinction between \allNFail\ and \mostNFail.
In order to detect 100\% of the failures, one shall consider robust patterns such as the dual self-checking pair pattern~\cite{963314}.
The constant \never\ denotes that the system shall never fail operational, fail silent or fail safe, \eg, the system shall never fail silent when a failure is detected.

\begin{example}
Consider the results of the safety analysis described in Section~\ref{sec:headlamp}.
A user may specify such results in our DSL as follows:

\begin{flushleft}
\small{
\code{hz(hz1,[hls,s3,e4,c3]). hz(hz2,[hls,s3,e3,c2]). ft(ft1,[bdCtl]). fl(fl1,[err]).}\\
\code{sg(sg1,[hz1,all1fail,never,all2fail]). sg(sg2,[hz2,never,most1fail,never]).}\\
\code{ft(ft2,[bcps]). ft2fl(ft1,fl1). fl(fl2,[loss]). ft2fl(ft2,fl2). ft(ft3,[cam]).}\\ 
\code{fl(fl3,[err]). ft2fl(ft3,fl3). mcs(mcs1,[fl1]). mcs(mcs2,[fl2]).}\\
\code{lmcs2hz([mcs1,mcs2],hz1). mcs(mcs3,[fl3]). lmcs2hz([mcs3],hz2).}}
\end{flushleft}
The facts \code{hz(hz1,[hls,s3,e4,c3])} and \code{hz(hz2,[hls,s3,e3,c2])} denote, respectively, the hazards \textbf{HZ1} and \textbf{HZ2} identified in Section~\ref{sec:headlamp}.
Consider hazard \code{hz1}.
The safety goal for addressing hazard \code{hz1} is specified by the fact \code{sg(sg1,[hz1,all1fail,never,all2fail])}.
The faults that trigger failures leading to \code{hz1} are specified as \code{ft(ft1,[bdCtl])} and \code{ft(ft2,[bcps])}.
These faults are associated, respectively, to the Body Control and to the logical channel between the Body Control and the Power Switch.
The fault \code{ft1} triggers failure \code{fl(fl1,[err])} of type erroneous, and the fault \code{ft2} triggers failure \code{fl(fl2,[loss])} of type loss.
The minimal cut set \code{mcs1} consists of failure \code{fl1} and \code{mcs2} consists of failure \code{fl2}.
Either minimal cut set \code{mcs1} or \code{mcs2} leads to hazard \code{hz1} (as specified by the fact \code{lmcs2hz([mcs1,mcs2],hz1)}).

\end{example}

Security-wise, we expect that a user (\eg, a security engineer) provides all public elements as input to \machinery.
The considered public elements for the headlamp system is shown in the example below.
We, however, do not expect a user to provide (potential) threats as input to \machinery, even though it is completely possible to do so. 
Instead, we derive (potential) threats from identified faults and failures.
The association of faults/failures and (potential) threats can be derived by reasoning rules, as demonstrated in Section~\ref{sec:security}.

\begin{example}
Consider the platform architecture of the headlamp system illustrated in Figure~\ref{fig:hls-architecture}.
Specified in our DSL, we consider the following hardware units as public:
\begin{flushleft}
\small{\code{public(int1). public(int2). public(int3).}}
\end{flushleft}
These facts denote, respectively, the Interfaces \code{int1}, \code{int2}, and \code{int3}.
\end{example}

\subsubsection{Safety and Security Patterns} Our DSL enables the specification of safety patterns for addressing failures, and security patterns for mitigating threats.
Table~\ref{table:dsl-safsecpat} describes the predicates for instantiating a pattern and for specifying the intent of a pattern.
The former includes the necessary components (\eg, the faulty component for safety) and channels for the pattern. 
The latter represents the intent of the pattern, including for which type of failure or threat the pattern is suitable to be applied.

Sections~\ref{sec:safety} and \ref{sec:security} describe how to declaratively specify safety and security patterns by example using the predicates \code{\safetyPattern}, \code{\safetyIntent}, \code{\securityPattern}, and \code{\securityIntent}.
\begin{table}
  \begin{tabular}{p{4cm}p{9cm}}
    \toprule
    \textbf{Fact} & \textbf{Description}\\
    \midrule
    \safetyPattern(\safetyPatternID,[\safetyPatternNAME,
    [\safetyPatternCP], [\safetyPatternINP],[\safetyPatternINT],[\safetyPatternOUT]]) & \safetyPatternNAME\ is a safety pattern of ID \safetyPatternID. This patter consists of a list of components (\eg, redundant components) \safetyPatternCP. The last three parameters \safetyPatternINP, \safetyPatternINT\ and \safetyPatternOUT\ denote, respectively, the input, the internal and the output channels related to the pattern.\\
    \midrule
    \safetyIntent(\safetyPatternNAME,[[\tp],
    \asil, \fop,\fsilent,\fsafe]) & \safetyPatternNAME\ is a safety pattern suitable for avoiding failures of type \tp, where \tp\ $\in$ \{\err,\loss\}. ASIL \asil\ denotes to which ASIL the pattern is suitable to be applied, where \asil\ $\in$ \{\asila,\asilb,\asilc,\asild\}. Pattern \safetyPatternNAME\ ensures the system to either fail operational (\fop), fail silent (\fsilent), or fail safe (\fsafe).\\  
   \midrule
    \securityPattern(\safetyPatternID,[\safetyPatternNAME,
    [\safetyPatternCP],[\safetyPatternINP],[\safetyPatternOUT]]) & \safetyPatternNAME\ is a security pattern of ID \safetyPatternID. This patter consists of a list of components \safetyPatternCP. The last three parameters \safetyPatternINP, \safetyPatternINT\ and \safetyPatternOUT\ denote, respectively, the input, the internal and the output channels related to the pattern.\\
    \midrule
    \securityIntent(\safetyPatternNAME,[[\trtype]]) & \safetyPatternNAME\ is a safety pattern suitable for mitigating threats of type \trtype, where $\trtype \in \{\confidentiality,\integrity,\availability\}$.\\  
    \bottomrule
  \end{tabular}
  \caption{\machinery: Language for safety and security architecture patterns}
  \label{table:dsl-safsecpat}
\end{table}

\section{Specification of Safety Architecture Patterns}\label{sec:safety}
This section illustrates how we can use \machinery to provide semantically-rich description of safety patterns that will enable the automated reasoning described in Section~\ref{sec:automation}. In particular, we instantiate the pattern template described in Section~\ref{sec:preliminaries} with two safety patterns.
For each instantiation, we provide a high-level description of the pattern and its specification in \machinery.
The pattern template includes pattern assumptions and security consequences from applying safety patterns.
The assumptions described in this section are not meant to be comprehensive.

\subsection{Dual Self-checking Pair with Fail Safe}\label{ssec:dual-self-checking-pair}

The dual self-checking pair pattern~\cite{963314} with fail safe consists of two pairs and a fault detector for each pair.
Each pair consists of a primary and a secondary component that are identical and operate in parallel. 
The primary and secondary components from the second pair are developed with a different design implementation in comparison to the components from the first pair.
The computations from each pair are sent to their respective fault detector.
While no failure is detected, the actuator receives the computations from the first pair.
The fault detector requires exact agreement from the computations (\ie, identical output values). 
When there is no exact agreement between the computations from the first pair, the fault detector of the first pair sends a take-over signal to the fault detector of the second pair.
This means that the computations from the second pair will be considered, and they will be sent to the actuator if the components produce identical output values.
If a failure is also detected on the second pair, the fault detector transition the system to a safe state.
This pattern fails operational always after the 1st failure of type erroneous (\ie, failure on the first pair).
It fails safe always after the 2nd failure is detected (\ie, failure on the second pair).
This pattern never fails silent.
The instantiation of the dual self-checking pair pattern with fail safe is shown in Table~\ref{table:dual-self-checking-pair-fs}. 

\begin{table}[t]
\centering
\begin{tabular}[t]{p{2cm}p{5.3cm}p{6cm}}
\toprule
&Description &\machinery Specification\\
\midrule
Pattern name&Dual self-checking pair pattern with fail safe&\code{NAME=dualSelfCheckingPairFS;}\\
\hline
Structure&\begin{minipage}{.3\textwidth}\includegraphics[scale=0.33]{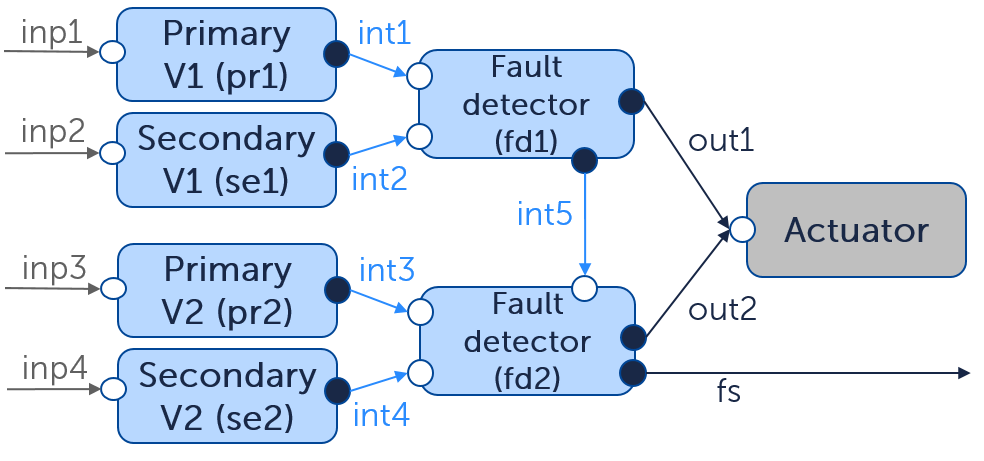}\end{minipage}
&
\code{\small{COMPONENT=[pr1,se2,fd1,pr2,se2,fd2];}}\newline 
\code{\small{INPUT\_CH=[inp1,inp2,inp3,inp4];}}\newline 
\code{\small{INTERNAL\_CH=[int1,int2,int3,int4,int5];}}\newline
\code{\small{OUTPUT\_CH=[out1,out2,fs];}}
\\
\hline
Intent&This pattern is suitable for high criticality hazards (ASIL C and D). This pattern fails operational always after the 1st erroneous failure, and it transition the system to a safe state always after the 2nd failure has been detected. This pattern never fails silent.&
\code{TYPE\_FAIL=[err];}\newline 
\code{ASIL=d;}\newline                                       
\code{FAIL\_OP=all1fail;}\newline  
\code{FAIL\_SILENT=never;}\newline 
\code{FAIL\_SAFE=all2fail;}
\\
Problem 
addressed&This pattern tolerates faults by avoiding failures of type erroneous.&\\
\hline
Assumptions&The first and second pair shall be implemented using independent designs. &\code{TYPE\_ASSUMPTION=are\_independent;}
\code{COMPONENT={[pr1,se1,pr2,se2]};}\\\cdashline{2-3}
&The the first and second pair shall be allocated to dedicated hardware units. &\code{TYPE\_ASSUMPTION=are\_decoupled;}
\code{COMPONENT={[pr1,se1,pr2,se2]};}\\\cdashline{2-3}
& The fault detector shall be verified. &\code{TYPE\_ASSUMPTION=are\_verified;}
\code{COMPONENT={[fd1,fd2]};}\\\cdashline{2-3}
\hline
Consequences (security)&There is a potential threat associated with the fault detectors. That is, as an intruder may carry out malicious actions to prevent the fault detectors from properly functioning. This potential threat is of type \textbf{int}egrity if the failure associated with the primary component is of type erroneous. &
\code{COMPONENT=[fd1,fd2];}
\code{TYPE\_THREAT=int;}
\\
\bottomrule
\end{tabular}
\caption{Dual self-checking pair pattern with fail safe}
\label{table:dual-self-checking-pair-fs}
\vspace*{-1cm}
\end{table}%


We describe the specification of the dual self-checking pair pattern with fail safe in \machinery.
Consider the language for safety patterns described in Table~\ref{table:dsl-safsecpat} and the structure of the pattern illustrated in Table~\ref{table:dual-self-checking-pair-fs}. 
This pattern is instantiated as follows:
\begin{center}
\code{safetyPattern(idpat,[dualSelfCheckingPairFS,[pr1,se1,fd1,pr2,se2,fd2],
[inp1,inp2,inp3,inp4],[int1,int2,int3,int4,int5],[out1,out2,fs]]).}
\end{center}

The safety intent of the dual self-checking pair pattern with fail safe is specified as follows:

\begin{center}
\code{safetyIntent(dualSelfCheckingPairFS,[[err],d,all1fail,never,all2fail]).}
\end{center}

Consider the assumptions for the dual self-checking pair pattern from Table~\ref{table:dual-self-checking-pair-fs}.
These assumptions will be created whenever this pattern is instantiated.
The first assumption in Table~\ref{table:dual-self-checking-pair-fs} is specified as follows in \machinery.

\begin{flushleft}
\code{assumption(dualSelfCheckingPairFS,are\_independent,[pr1,se1,pr2,se2])}\\
\hspace{0.2cm}\code{:- safetyPattern(idpat,[dualSelfCheckingPairFS,[pr1,se1,\_,pr2,se2,\_],\_,\_,\_]).}
\end{flushleft}

\subsection{Monitor-Actuator Pattern}\label{ssec:monitor-actuator}

The monitor-actuator pattern~\cite{Armoush2010} consists of a primary component and a monitor.
The monitor consumes the computations from the primary component and its original inputs such that it can cross-check their computations to identify failures of type erroneous.
While no failure is detected by the monitor (\eg, through the use of plausibility checks), the actuator receives the outputs from the primary component.
If the monitor detects a failure on the primary component, the monitor initiates a corrective action by sending a shutdown signal to the primary component. 
That is, this pattern fails silent always after most 1st failure of type erroneous have been detected on the primary component.
It neither fails operational nor fails safe.
The instantiation of the monitor-actuator pattern is shown in Table~\ref{table:monitor-actuator}. 

\begin{table}[t]
\centering
\begin{tabular}[t]{p{2cm}p{5.5cm}p{5.5cm}}
\toprule
&Description &\machinery Specification\\
\midrule
Pattern name&Monitor-Actuator Pattern&\code{NAME=monitorActuator;}\\
\hline
Structure&\begin{minipage}{.3\textwidth}\includegraphics[scale=0.35]{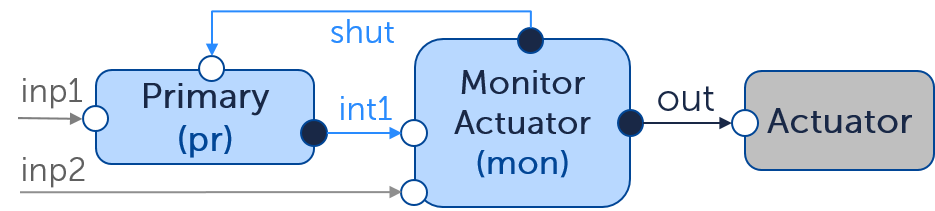}\end{minipage}
&
\code{COMPONENT=[pr,mon];} 
\code{INPUT\_CH=[inp1,inp2];} 
\code{INTERNAL\_CH=[int1,shut];}
\code{OUTPUT\_CH=[out];}
\\
\hline
Intent&This pattern is suitable for low criticality hazards (ASIL A and B). This pattern fails silent always after most 1st failures. It never fails operational and it never fails safe.&
\code{TYPE\_FAIL=[err];}\newline  
\code{ASIL=b;}\newline                                       
\code{FAIL\_OP=never;}\newline  
\code{FAIL\_SILENT=all1fail;}\newline 
\code{FAIL\_SAFE=never;}
\\
Problem addressed&This pattern tolerates faults by avoiding failures of type erroneous.&\\
\hline
Assumptions&The monitor shall be verified. &\code{TYPE\_ASSUMPTION=are\_verified;}
\code{COMPONENT={[mon]};}\\
\hline
Consequences (security)&There is a potential threat associated with the monitor, as an intruder may carry out malicious actions to prevent the monitor from properly functioning. This potential threat is of type \textbf{int}egrity if the failure associated with the primary component is of type erroneous. &
\code{COMPONENT=[mon];}
\code{TYPE\_THREAT=int;}
\\
\bottomrule
\end{tabular}
\caption{Monitor-Actuator pattern}
\label{table:monitor-actuator}
\vspace*{-0.8cm}
\end{table}%

We describe the specification of the monitor-actuator pattern in \machinery.
Consider the language for safety patterns described in Table~\ref{table:dsl-safsecpat} and the structure of the pattern illustrated in Table~\ref{table:monitor-actuator}. 
This pattern is instantiated as follows:
\begin{flushleft}
\code{safetyPattern(idpat,[monitorActuator,[pr,mon],[inp1,inp2],[int1,shut],[out]]).}
\end{flushleft}

The safety intent of the monitor-actuator pattern is specified as follows:

\begin{flushleft}
\code{safetyIntent(monitorActuator,[[err],b,never,most1fail,never]).}
\end{flushleft}

We specify the assumption for the monitor-actuator pattern from Table~\ref{table:monitor-actuator} as follows.
This assumption will be created whenever a monitor-actuator pattern is instantiated.

\begin{flushleft}
\code{assumption(monitorActuator,are\_verified,[mon])}\\
\hspace{0.2cm}\code{:- safetyPattern(idpat,[monitorActuator,[\_,mon],\_,\_,\_]).}
\end{flushleft}


\section{Specification of Security Architecture Patterns}\label{sec:security}
As in the previous section, this section illustrates how we can use \machinery to provide semantically-rich description of 
security patterns that will enable the automated reasoning described in Section~\ref{sec:automation}.
We instantiate the pattern template described in Section~\ref{sec:preliminaries} with two security patterns.
For each instantiation, we provide a high-level description of the pattern and its specification in \machinery.
The pattern template includes pattern assumptions and safety consequences from applying the security pattern.
The assumptions are not meant to be comprehensive.

\subsection{Firewall Pattern}\label{ssec:firewall}

\begin{table}[t]
\centering
\begin{tabular}[t]{p{2cm}p{5.5cm}p{5.5cm}}
\toprule
&Description &\machinery Specification\\
\midrule
Pattern name&Firewall&
\code{NAME=firewall;}\\
\hline
Structure&\begin{minipage}{.3\textwidth}\includegraphics[scale=0.3]{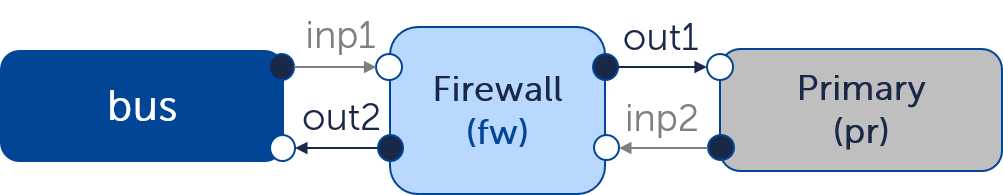}\end{minipage}&
\code{COMPONENT=[bus,fw,pr];} 
\code{INPUT\_CH=[inp1,inp2];} 
\code{OUTPUT\_CH=[out1,out2];}
\\
\hline
Intent&This pattern intercepts, filters, and blocks incoming and outgoing messages in the network layer.&\code{TYPE\_THREAT=[ava,int];} 
\\\cline{1-2}
Problem addressed&This pattern mitigates threats that violates the availability and integrity of the system.&\\
\hline
Assumptions&Firewall shall be verified &\code{TYPE\_ASSUMPTION=are\_verified;}
\code{COMPONENT={[fw]};}\\\cdashline{2-3}
&Security policies shall be specified. &\code{TYPE\_ASSUMPTION=have\_policies;}
\code{COMPONENT={[fw]};}\\\cdashline{2-3}
\hline
Consequences (safety)&The deployed firewall might be faulty, \eg, it might erroneously block legit messages. Thus, there is a new fault associated with the deployed firewall that may trigger erroneous failures.&
\code{COMPONENT=[fw];}
\code{TYPE\_FAILURE=err;}
\\
\bottomrule
\end{tabular}
\caption{Firewall pattern}
\label{table:firewall}
\end{table}

The firewall pattern~\cite{SljivoUPG20} is instantiated in Table~\ref{table:firewall}. 
A firewall is placed between a bus (\eg, a CAN bus) and a component. 
The bus receives and sends messages from the external and internal network, respectively.
These messages are intercepted and analyzed by the firewall.
The firewall mitigates threats of type availability and integrity.
That is, the firewall controls the network access to the internal network according to predefined security policies (\eg, blacklisting IP addresses consuming more bandwidth than a given threshold), and can also inspect message content to detect intrusion attempts and anomalies~\cite{SljivoUPG20}.  

We describe the specification of the firewall pattern in \machinery.
Consider the language for security patterns described in Table~\ref{table:dsl-safsecpat} and the structure of the pattern illustrated in Table~\ref{table:firewall}. 
The firewall pattern is instantiated as follows:
\begin{center}
\code{securityPattern(idpat,[firewall,[bus,pr,fw],[inp1,inp2],\_,[out1,out2]]).}
\end{center}

The security intent of the firewall pattern is specified as follows:

\begin{center}
\code{securityIntent(firewall,[[ava,int]]).}
\end{center}

The first assumption for the firewall pattern from Table~\ref{table:firewall} is specified as follows.

\begin{flushleft}
\code{assumption(firewall,are\_verified,[fw])}\\
\hspace{0.2cm}\code{:- securityPattern(idpat,[firewall,[\_,\_,fw],\_,\_,\_]).}
\end{flushleft}

\subsection{Security Monitor Pattern}\label{ssec:security-monitor}

\begin{table}[t]
\centering
\begin{tabular}[t]{p{2cm}p{5.5cm}p{5.5cm}}
\toprule
&Description &\machinery Specification\\
\midrule
Pattern name&Security Monitor&
\code{NAME=securityMonitor;}\\
\hline
Structure&\begin{minipage}{.3\textwidth}\includegraphics[scale=0.3]{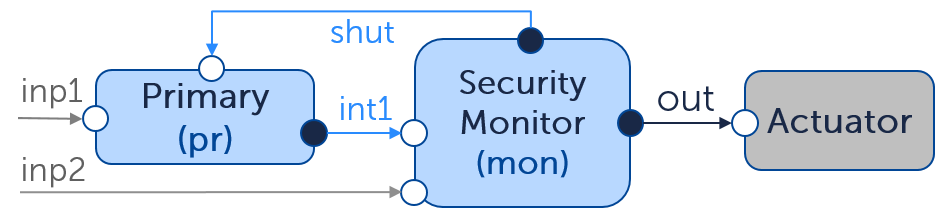}\end{minipage}&
\code{COMPONENT=[pr,mon];} 
\code{INPUT\_CH=[int1,int2];} 
\code{INTERNAL\_CH=[int1,shut];} 
\code{OUTPUT\_CH=[out];}
\\
\hline
Intent&This pattern intercepts, filters, and blocks incoming and outgoing messages from components.&\code{TYPE\_THREAT=[int];} 
\\\cline{1-2}
Problem addressed&This pattern mitigates threats that violates the integrity of the system.&\\
\hline
Assumptions&Firewall shall be verified &\code{TYPE\_ASSUMPTION=are\_verified;}
\code{COMPONENT={[mon]};}\\\cdashline{2-3}
&Security policies shall be specified. &\code{TYPE\_ASSUMPTION=have\_policies;}
\code{COMPONENT={[mon]};}\\\cdashline{2-3}
\hline
Consequences (safety)&The deployed monitor might be faulty, \eg, it might erroneously block legit messages. Therefore, there is a new fault associated with the deployed monitor that may trigger erroneous failures.&
\code{COMPONENT=[mon];}
\code{TYPE\_FAILURE=err;}
\\
\bottomrule
\end{tabular}
\caption{Security monitor pattern}
\label{table:security-monitor}
\end{table}

The security monitor pattern is instantiated in Table~\ref{table:security-monitor}. 
This pattern mitigates threats that violate the integrity of the system.
We consider security monitors that mitigate such threats by monitoring incoming and outgoing messages from components and enforcing security policies in the application layer.
Whenever a security policy is violated, the monitor can initiate a corrective action by sending a shutdown signal to the component~\cite{GayMS11}. 

We describe the specification of the security monitor pattern in \machinery.
Consider the language for security patterns described in Table~\ref{table:dsl-safsecpat} and the structure of the pattern illustrated in Table~\ref{table:security-monitor}. 
The security monitor pattern is instantiated as follows:
\begin{center}
\code{securityPattern(idpat,[securityMonitor,[pr,mon],[inp1,inp2],[int1,shut],[out]]).}
\end{center}

The security intent of the security monitor pattern is specified as follows:

\begin{center}
\code{securityIntent(securityMonitor,[[int]]).}
\end{center}

The second assumption for the security monitor described in Table~\ref{table:security-monitor} is specified as follows.

\begin{flushleft}
\code{assumption(securityMonitor,have\_policies,[mon])}\\
\hspace{0.2cm}\code{:- securityPattern(idpat,[securityMonitor,[\_,mon],\_,\_,\_]).}
\end{flushleft}
\section{Safety and Security Reasoning Principles}\label{sec:automation}

Building on the DSL presented in Section~\ref{sec:framework}, this section describes some safety and security reasoning principles that can be automated by using solvers, such as DLV~\cite{dlv}.

\subsection{Safety Reasoning}\label{ssec:automation-safety-patterns}
Building on top of~\cite{dantas20iclp}, we specify as logic programs safety reasoning principles to determine when (a) a failure can be avoided, (b) a minimal cut set can be avoided, (c) a fault can be tolerated, (d) a hazard can be controlled, and (e) a safety goal can be satisfied.
We introduce five new facts to specify safety reasoning principles for (a), (b), (c), (d) and (e). 
Note that four of the new facts receive attributes from the safety pattern intent as argument in order to make explicit how they have been addressed by the pattern.
These attributes consists of ASIL values (\eg, \code{a}), fail operational values (\eg, \code{all1fail}), fail silent values (\eg, \code{all2fail}), and fail safe values (\eg, \code{never}). 

\begin{itemize}
	\item \code{avoided(IDFL,ATTRSINTENT)} denotes that failure \code{IDFL} is avoided with a safety pattern intent \code{ATTRSINTENT} (\eg, avoided by a pattern that fails silent always after the 1st failure).
	\item \code{avoidedMCS(IDMCS,ATTRSINTENT)} denotes that minimal cut set \code{IDMCS} is avoided with a safety pattern intent \code{ATTRSINTENT}.
	\item \code{tol(IDFT,ATTRSINTENT)} denotes that fault \code{IDFT} is tolerated with a safety pattern intent \code{ATTRSINTENT}.
	\item \code{ctl(IDHZ,ATTRSINTENT)} denotes that hazard \code{IDHZ} is controlled with a safety pattern intent \code{ATTRSINTENT}.
	\item \code{satisfied(IDSG)} denotes that safety goal \code{IDSG} is satisfied.
\end{itemize}	

Specified by the next rule, a failure is avoided if a pattern is associated to the faulty component \code{TARGET}, and the pattern is able to avoid failures of type \code{TYPE} checked by \code{\#member(TYPE,PATTYPE)}. 

\begin{flushleft}
\small{
\code{avoided(IDFL,[IDPAT | ATTRSINTENT]) :- fl(IDFL,[TYPE]),}\\
\hspace{1cm}\code{ft(IDFT,[TARGET]), ft2fl(IDFT,IDFL), \#member(TYPE,PATTYPE),}\\
\hspace{1cm}\code{getSafPatTarget(IDPAT,TARGET),safetyIntent(PAT,[PATTYPE | ATTRSINTENT]),}\\			                 
\hspace{1cm}\code{safetyPattern(IDPAT,[PAT | ATTRSPAT]).}}
\end{flushleft}

A minimal cut set \code{IDMCS} is avoided if at least one failure of its set has been avoided. 
A fault is tolerated if the failures triggered by that fault are avoided.
Both the rule for \code{tol} and for \code{avoidedMCS} are omitted here.
Next, we specify a rule for hazard controllability.

\begin{flushleft}
\small{\code{ctl(IDHZ,ATTRSINTENT) :- hz(IDHZ,ATTRSHZ),}\\
\hspace{1cm}\code{lmcs2hz(LMCS,IDHZ), getMinIntent(LMCS,ATTRSINTENT).}}
\end{flushleft}

A hazard is controlled if each MCS in the list of \code{LMCS} is avoided.
This is checked by the fact \code{getMinIntent(LMCS,ATTRSINTENT)}. 
In addition, \code{getMinIntent(LMCS,ATTRSINTENT)} returns the minimal attributes needed for controlling hazard \code{IDHZ} (see example below). 
This is relevant to show the minimal attributes (taken from the pattern intent) required to control the hazard.

\begin{example}\label{ex:getMinIntent}
Consider two failures \code{FLA} and \code{FLB} that leads to hazard \code{HZA}.
Safety pattern \code{SPA} avoids failure \code{FLA} with the intent attributes \code{[c,all1fail,never,all2fail]}, and safety pattern \code{SPB} avoids failure \code{FLB} with the intent attributes \code{[d,all1fail,never,never]}. 
The fact \code{getMinIntent(LMCS,ATTRSINTENT)} will return the minimal attributes for controlling hazard \code{HZA}, \ie, \code{[c,all1fail,never,never]}.
\end{example}

The next rule denotes when a safety goal is satisfied.
A safety goal \code{IDSG} is satisfied if hazard \code{IDHZ} is controlled with higher or equal attributes than the ones required by the safety goal (similarly to Example~\ref{ex:getMinIntent}).
These checks are done by the fact \code{checkHigherOrEqual(IDHZ,IDSG)}.

\begin{flushleft}
\small{\code{satisfiedSG(IDSG) :- sg(IDSG,[IDHZ | ATTRSSG),}\\
\hspace{0.5cm}\code{getHazardASIL(IDHZ,ASIL), ctl(IDHZ,ATTRSCTL), checkHigherOrEqual(IDHZ,IDSG).}}
\end{flushleft}

\subsubsection{Recommendation of Safety Patterns}
We now introduce our reasoning rule for recommending which safety patterns could be used at which place of the system architecture to avoid failures provided as input information by the user.
This is done by using the following disjunctive rule:
\begin{flushleft}
\small{\code{xsafetyPattern([nuIDSAFPAT,PAT,CTR],[PAT,[TARGET,[nuRED,CTR],[nuCKR,CTR]],}\\
\code{[nuINP,CTR],[nuINT,CTR],[nuOUT,CTR]]) v }\\
\code{nxsafetyPattern([nuIDSAFPAT,PAT,CTR],[PAT,[TARGET,[nuRED,CTR],[nuCKR,CTR]],}\\
\code{[nuINP,CTR],[nuINT,CTR],[nuOUT,CTR]])}\\
\hspace{1cm}\code{:- fl(IDFL,[FLTYPE]), ft(IDFT,[TARGET]), ft2fl(IDFT,IDFL),}\\
\hspace{1cm}\code{exploreSafPat(PAT), safetyIntent(PAT,[PATFLTYPE | ATTRSINTENT]),}\\
\hspace{1cm}\code{\#member(FLTYPE,PATFLTYPE), getSafIntentASIL(PAT,PATASIL), mcs(IDMCS,FAILURES),}\\
\hspace{1cm}\code{\#member(IDFL,FAILURES), lmcs2hz(IDMCS,IDHZ), asil(IDHZ,HZASIL),}\\
\hspace{1cm}\code{higherEqualThan(PATASIL,HZASIL), counterSafPat(CTR).}}
\end{flushleft}


It specifies the recommendation (\code{xsafetyPattern}) or not (\code{xnsafetyPattern}) of a safety pattern.
Specifically, this rule specifies that a safety pattern is recommended to avoid a failure \code{IDFL} of type \code{FLTYPE} triggered by fault \code{IDFT} associated to component \code{TARGET} that lead to hazard \code{IDHZ} if the safety pattern is suitable for both avoiding \code{FLTYPE} as checked by \code{\#member(TYPE,PATTYPE)} and addressing the ASIL of \code{IDHZ} as checked by \code{higherEqualThan(PATASIL,HZASIL)}. \code{PATTYPE} is taken from the pattern intent.

Note that the prefix ``\code{x}'' in front of \code{safetyPattern} is to make explicit that the safety pattern has been automatically recommend by \machinery.
Omitted here, we have a rule for mapping \code{xsafetyPattern} to \code{safetyPattern}.
The fact \code{exploreSafPat(PAT)} denotes that \machinery shall explore whether the safety pattern \code{PAT} is suitable to avoid a given failure.
Which patterns shall be considered by \machinery is provided by the user.
The fact \code{counterSafPat(CTR)} denotes a counter \code{CTR} to ensure that each safety pattern has a unique ID.
The constants starting with \code{nu} do not appear in the baseline architecture. 
Whenever a safety pattern is recommended, \machinery ensures that both the components and channels related to the recommended pattern are created. 
These components and channels are prefixed with \code{nu} so that one can easily identify the increments in the architecture modified by \machinery.

As the system complexity grows and with it the number of failures and 
locations where a safety pattern can be placed, the number of pattern recommendations may rapidly increase.
To keep the number of models manageable, we use DLV constraints to \emph{not consider models} where, \eg, the same instance of a pattern is recommended more than once and more than one suitable pattern is recommended to avoid a given failure (\ie, to avoid that 2 distinct patterns avoid the same failure).

\subsubsection{Safety Pattern Recommendation for the Headlamp System}

We now apply \machinery to the headlamp system described in Section~\ref{sec:headlamp} to automatically select which safety patterns could be used to avoid the identified failures.
We run our recommendation machinery explained in the section above for a number of safety patterns, including the dual self-checking pair pattern, the heterogeneous duplex pattern~\cite{Armoush2010}, the monitor-actuator pattern, and the watchdog pattern~\cite{Armoush2010}.

\begin{wrapfigure}{r}{0.45\textwidth}
  \vspace*{-7mm}
  \begin{center}
    \includegraphics[width=0.45\textwidth]{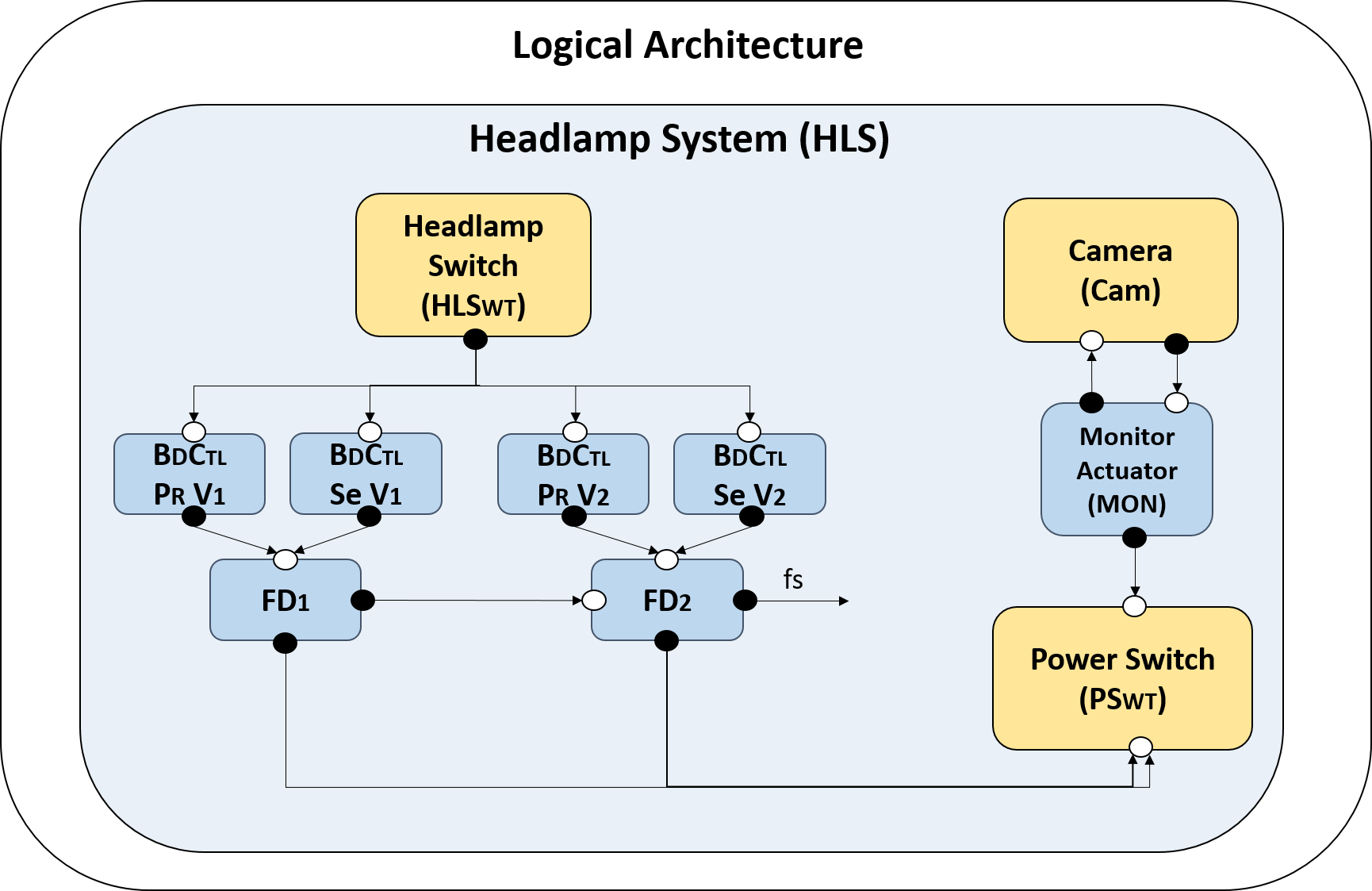}
  \end{center}
  \caption{Headlamp system with patterns}
  \label{fig:hls-safety-patterns}
\end{wrapfigure}

We consider the defined safety goals \code{SG1} and \code{SG2} to address hazards \code{HZ1} and \code{HZ2}, respectively.
The goal is to provide safety patterns suitable for (a) avoiding the failures leading to hazards \code{HZ1} and \code{HZ2}, and (b) satisfying the safety goals. 
Figure~\ref{fig:hls-safety-patterns} illustrates one architecture solution provided by \machinery that achieves this goal.
Note that Figure~\ref{fig:hls-safety-patterns} omits the external components from the headlamp system due to the lack of space. 

\machinery recommended the dual self-checking pair pattern for avoiding erroneous failures (\ie, failure \code{FL1}) on the Body Control.
This pattern satisfies the safety goal \code{SG1}, as it fails operational always after the 1st failure, and it fails safe always after the 2nd failures.
\machinery recommended the monitor-actuator pattern for avoiding erroneous failures (\ie, failure \code{FL3}) on the Camera.
This pattern satisfies the safety goal \code{SG2}, as it fails silent always after the 1st failure.
The pattern assumptions generated by \machinery can help engineers to deploy the pattern-related components into the platform architecture.

\subsection{Security Reasoning}\label{ssec:automation-security-patterns}

This paper proposes the use of KRR for security with architecture patterns. 
The goal is to provide automated methods for automating the recommendation of security architecture patterns to mitigate threats.
As a basis to achieve this goal, we specify security reasoning principles to determine when (a) a potential threat becomes a threat, and (b) a threat can be mitigated by a security pattern. 

Before introducing these reasoning principles, we first introduce our threat model.

\subsubsection{Threat Model}
The threat model we assume is inspired by the traditional Dolev-Yao intruder model~\cite{DY} widely used for security protocol verification. Intuitively, the DY intruder is the most powerful symbolic intruder. She can have access and manipulate any information to which she has access to, \ie, information that appears in a channel reachable from a public interface that is not encrypted or encrypted with a key that she possesses the decryption key. 

More precisely, our intruder model has the following capabilities inspired by the Dolev-Yao model for CAN bus communication channels:
\begin{itemize}
  \item \textbf{Base Case:} The intruder can reach any hardware that is deployed in a public hardware/interface. 
  \item \textbf{Inductive Case 1:} If the intruder can reach hardware \code{HW} and there is a component \code{CP} deployed in \code{HW} that writes on a CAN bus \code{CAN}, then the intruder can also reach \code{CAN};
  \item \textbf{Inductive Case 2:} If the intruder can reach the CAN Bus \code{CAN}, and there is a component \code{CP} deployed in hardware \code{HW} that reads from the \code{CAN}, then the intruder can reach \code{HW}. 
\end{itemize}

\begin{figure}
  \begin{center}
    \includegraphics[width=11cm]{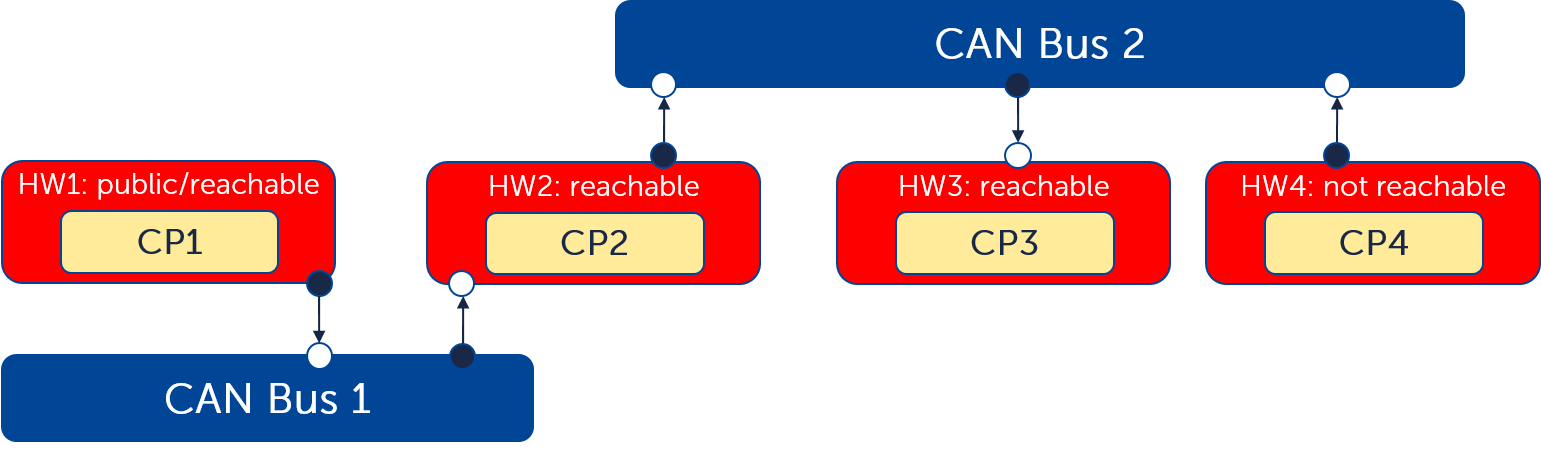}
  \end{center}
  \caption{Illustration of the intruder reachability}
  \label{fig:intruder-model-example}
  \end{figure}

For example, consider the topology depicted in Figure~\ref{fig:intruder-model-example}. 
Assume that the hardware unit \code{HW1} is public, \eg, it has a wireless interface. Therefore, the intruder can reach to \code{HW1}. Furthermore, assume that a component, \code{CP1}, deployed in \code{HW1} writes to CAN bus \code{CAN Bus 1}, then from the inductive case 1, the intruder can reach \code{CAN Bus 1}, \ie, she can (in principle) write into \code{CAN Bus 1}. 
Furthermore, assume that a component \code{CP2} deployed in hardware \code{HW2} reads from \code{CAN Bus 2}, then from inductive case 2, the intruder can also reach \code{HW2}.
Similarly, since there is a component in \code{HW2} that writes into \code{CAN Bus 2} and a component in \code{HW3} that reads from \code{CAN Bus 2}, the intruder can reach \code{CAN Bus 3} and \code{HW3}. 
However, since there is no component in \code{HW4} that reads from \code{CAN Bus 2}, but possibly only writes into \code{CAN Bus 2}, the intruder cannot reach \code{HW4}.

We can easily program/customize this intruder model as logic programming. 
For example, the intruder model described above is specified by a number of rules in \machinery, the rules for the \textbf{Base Case} and \textbf{Inductive Case 1} are, respectively, shown below, while the case for \textbf{Inductive Case 2} follows similarly and is elided:

\begin{flushleft}
\small{
\code{reachI(CP,HWCP,[HWCP]) :- public(CP), dep(CP,HWCP).}\\

\code{reachI(CH,CAN,[CAN|PATH]) :- writesToCan(CP,CAN), dep(CH,CAN),} \\
\hspace{1cm}\code{dep(CP,HWCP), reachI(CP,HWCP,PATH), not \#member(CAN,PATH).}}
\end{flushleft}


\noindent\textbf{Remarks:} Notice that while we are inspired by the Dolev-Yao intruder model, there is a key difference. Since we are not taking into account the contents of the messages exchanged, the intruder model shown above does not take into account the fact that messages are encrypted or not nor does it take into account on how the messages exchanged are actually used. 
This means that the reachability to components and communication channels is an over-approximation which may lead to false positives.
To make the the analysis more precise, one could include more information about the messages exchanged, \eg, whether a message is encrypted. 
This is, however, left out of the scope of this paper as it deals with different phases of development.

We also notice that the threat model could be further refined by considering vulnerabilities as done in~\cite{attack-paths}. This would mean the extension of our DSL with vulnerabilities and would, in principle, enable a 
more refined analysis of the possible attacks. Indeed, we believe that it is possible to specify logic programs 
encoding the rules described in~\cite{attack-paths}. We leave this exercise to future work.  

\subsubsection{Security for Safety}

To demonstrate the safety of the system, security engineers shall ensure that intruders cannot trigger identified failures. 
Hence, we specified security reasoning principles for deriving potential threats from identified failures.

Inspired by~\cite{durrwang17safecomp,durrwang18sae,iso21434}, we derive a potential threat from a failure based on the type and on severity the hazard led by the failure.
A failure of type \code{err} that leads to a hazard of severity \code{S3} is mapped to a potential threat (\texttt{pThreat}) of type \code{int} and of severity \code{sev}.
A failure of type \code{loss} that leads to a hazard of severity \code{S2} is mapped to a \texttt{pThreat} of type \code{ava} and of severity \code{maj}.
A failure of type \code{err} that leads to a hazard of severity \code{S1} is mapped to a \texttt{pThreat} of type \code{int} and of severity \code{mod}.
Currently, we are not considering a mapping from a failure type to confidentiality.
The mapping from failure to potential threat is specified by the next rule.


\begin{flushleft}
\small{\code{pThreat(IDFL,[TARGET,HWTARGET,SECTYPE,SECSEV]) :- fl(IDFL,[SAFTYPE]), ft(IDFT,[TARGET]),}\\ 
\hspace{0.3cm}\code{ft2fl(IDFT,IDFL), dep(TARGET,HWTARGET), typeMap(SAFTYPE,SECTYPE),}\\
\hspace{0.3cm}\code{hz(IDHZ,[\_,SAFSEV,\_,\_]), mcs(IDMCS,FLISTFAIL), \#member(IDFL,FLISTFAIL),}\\
\hspace{0.3cm}\code{lmcs2hz(LIDMCS,IDHZ), \#member(IDMCS,LIDMCS), severityMap(SAFSEV,SECSEV).}}
\end{flushleft}

Specified by the following rule, a potential threat becomes a threat if the hardware unit \textbf{HWTARGET} can be reached through a path \code{PATH} (as described in the threat model).

\begin{flushleft}
\small{\code{threat([IDPT,PATH],[TARGET,HWTARGET,SECTYPE,SECSEV]) :-}\\ 
\hspace{0.3cm}\code{pThreat(IDPT,[TARGET,HWTARGET,SECTYPE,SECSEV]), reachI(TARGET,HWTARGET,PATH).}}
\end{flushleft}

\begin{example}
Consider the failure \code{fl1} and hazard \code{hz1} identified in Section~\ref{sec:headlamp}.
Failure \code{fl1} that leads to hazard \code{hz1} is mapped to the potential threat \code{pt1}. 
The potential threats becomes a threat as \code{ecu2} can be reached by three paths, including the path from the Bluetooth (deployed into interface \code{int3}) to the Body Control (deployed into ECU \code{ecu2}).
\begin{center}
\small{\code{pThreat(pt1,[bdCtl,ecu2,err,sev]).}\\
\code{threat([pt1,[ecu2,can2,ecu3,can1,ecu4,int3],[bdCtl,ecu2,err,sev]).}}
\end{center}
The intruder path \code{[ecu2,can2,ecu3,can1,ecu4,int3]} can be read from right to left.
\end{example}

We introduce one new fact, namely \code{mit(IDTH)}, for when a threat \code{IDTH} is mitigated.
We omit the rule for \code{mit} here. 
In a nutshell, a threat \code{IDTH} is mitigated if a suitable security pattern for mitigating the type of threat violated by \code{IDTH} is placed in the architecture.

\subsubsection{Recommendation of Security Patterns}
We introduce our reasoning rule for recommending which security patterns could be used at which place of the system architecture to mitigate threats provided as input information by the user or derived by safety elements.
Our rules for recommending security patterns are tailored to the pattern.
For example, the firewall and the security monitor patterns are applied to distinct places in the system architecture. 
That is, the firewall is placed between a CAN bus and a hardware unit, and the security monitor is placed to an individual hardware unit.\footnote{Security patterns for automotive is an ongoing research topic~\cite{ChengDPP19}. In principle, we can also specify reasoning principles for security patterns that ensure confidentiality such as the Symmetric encryption pattern. For patterns that require encryption \machinery would not make any visible changes in the system architecture. Instead, \machinery would provide security requirements such as ``Channel X shall be encrypted''.}
The next rule specifies the placement (\code{xsecurityPattern}) or not (\code{nxsecurityPattern}) of the firewall pattern.


\begin{flushleft}
\small{
\code{xsecurityPattern([nuIDSECPAT,firewall,CTR],[firewall,[HWUNIT,COMM,[nuCKR,CTR]],}\\
\code{[nuINP,CTR],[nuINT,CTR],[nuOUT,CTR]]) v}\\
\code{xnsecurityPattern([nuIDSECPAT,firewall,CTR],[firewall,[HWUNIT,COMM,[nuCKR,CTR]],
[nuINP,CTR],[nuINT,CTR],[nuOUT,CTR]])}\\
\hspace{0.5cm}\code{:- threat(IDTR,ATTRSTR), getThreatType(IDTR,TRTYPE), getThreatTarget(IDTR,TARGET),}\\
\hspace{0.5cm}\code{getThreatPath(IDTR,PATH), can(COMM), hw(HWUNIT),\#subList([HWUNIT,COMM],PATH),}\\
\hspace{0.5cm}\code{exploreSecPat(firewall), securityIntent(firewall,ATTRSINTENT),counterSecPat(CTR),}\\
\hspace{0.5cm}\code{getSecIntentThreatType(firewall,PATTRTYPE), \#member(TRTYPE,PATTRTYPE).}}   
\end{flushleft}

This rule specifies that the firewall pattern is recommended to mitigate a threat \code{IDTR} exploited through path \code{PATH} if a firewall is placed between a can \code{COMM} and a hardware unit \code{HWUNIT}, where both \code{COMM} and \code{HWUNIT} are in \code{PATH} as checked by \code{\#subList([HWUNIT,COMM],PATH)}.
The firewall shall be able to mitigate the type of threat violated by \code{IDT} as checked by \code{\#member(TRTYPE,PATTRTYPE)}.
We leave to future work the use of severity of threats as a condition to recommend security patterns.

\subsubsection{Security Pattern Recommendation for the Headlamp System}
We now apply \machinery to the headlamp system described in Section~\ref{sec:headlamp} to automatically select which security patterns could be used to mitigate the identified threats.
We run our recommendation machinery for the firewall pattern and the security monitor pattern.

We consider the six threats derived from the identified failures on the headlamp system.
These threats target either the ECU \code{ecu2} (\ie, Body Control) or the CAN \code{can2} (\ie, logical communication between the Body Control and the Power Switch) from the public elements, that is, \code{int1} (\ie, OBD-II C.), \code{int2} (\ie, Cellular), \code{int3} (\ie, Bluetooth).
Two of these threats are shown below.
\begin{flushleft}
\small{
\code{threat([pt1,[ecu2,can2,ecu3,can1,ecu4,int3]],[bdCtl,ecu2,int,sev]).}\\ 
\code{threat([pt2,[can2,ecu3,can3,int1]],[bcps,can2,ava,sev]).}
}
\end{flushleft}

\begin{figure}
\begin{center}
	\includegraphics[width=\textwidth]{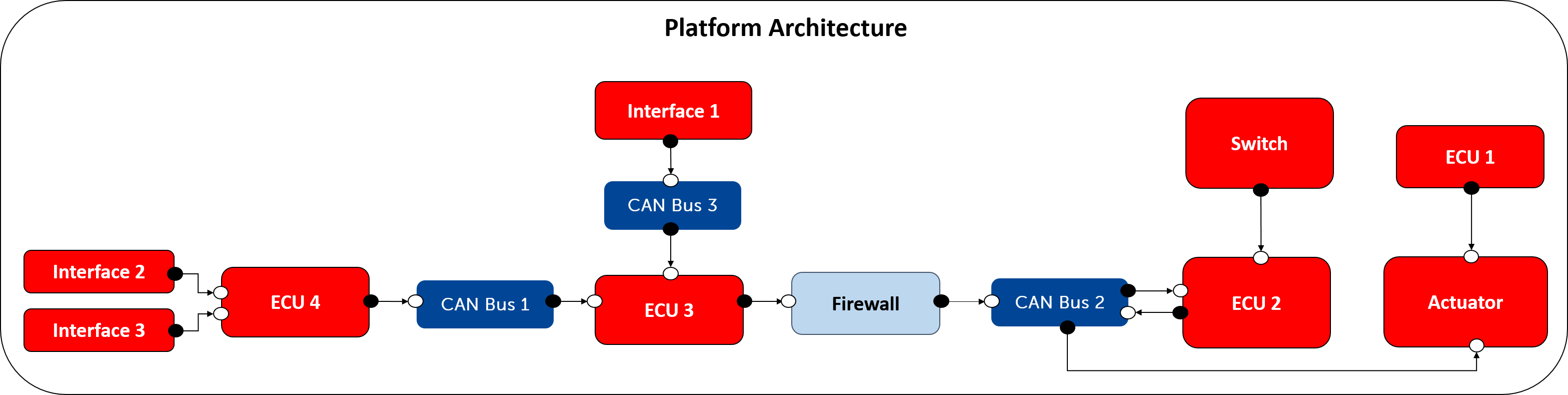}	
\end{center}
\caption{Headlamp system with security pattern.}
\label{fig:hls-security-patterns}
\end{figure}

Our goal is to provide security patterns suitable for mitigating these threats that violate the availability (\code{ava}) and the integrity (\code{int}) of the headlamp system.
Figure~\ref{fig:hls-security-patterns} illustrates one architecture solution provided by \machinery that achieves this goal.
\machinery recommended the firewall pattern for mitigating all derived threats.
The firewall is placed between \code{ecu3} and \code{can2} so that it can intercept, filter, and block incoming messages (possibly malicious) from public elements. 

\subsection{Safety and Security Co-Analysis Reasoning}\label{ssec:automation-safety-security-patterns}

This paper proposes the use of KRR for safety and security co-analysis with architecture patters.
The goal is to provide automated methods to reason about the consequences of safety patterns to security, and of security patterns to safety.

\subsubsection{Security Consequences Caused by Safety Patterns} 
We specified reasoning principles for determining when a security pattern can cause consequences to security.

The deployment of a safety pattern may lead to a new (potential) threat to the system, as an intruder may perform malicious actions to prevent the deployed safety pattern from properly functioning (\eg, not avoiding failures).
The following reasoning rule specifies this consequence. 

\begin{flushleft}
\small{
\code{pThreat(IDPAT,[CP,CHECKER,SECTYPE,SECSEV]) :- safetyPattern(IDPAT,ATTRSPAT),}\\
\hspace{0.5cm}\code{getSafPatChecker(IDPAT,CHECKER), getSafPatTarget(IDPAT,TARGET), ft(IDFT,[TARGET]),}\\
\hspace{0.5cm}\code{fl(IDFL,[FAILTYPE]), ft2fl(IDFT,IDFL), dep(CP,CHECKER), lmcs2hz(LIDMCS,IDHZ),}\\
\hspace{0.5cm}\code{typeMap(FAILTYPE,SECTYPE), hz(IDHZ,[\_,SAFSEV,\_,\_]), mcs(IDMCS,FLISTFAIL),}\\							               
\hspace{0.5cm}\code{\#member(IDFL,FLISTFAIL), \#member(IDMCS,LIDMCS), severityMap(SAFSEV,SECSEV).}							               
}
\end{flushleft}

There is a new potential threat \code{IDPAT} (same id of the safety pattern) associated with the \code{CHECKER} of the safety pattern (\eg, a fault detector) if \code{CHECKER} is monitoring a faulty component \code{TARGET}.
This potential threat becomes an actual threat if \code{CHECKER} can be reached by an intruder.

\subsubsection{Security Consequences on the Headlamp System}
Consider the headlamp system with safety patterns illustrated in Figure~\ref{fig:hls-safety-patterns}.
The deployment of a monitor-actuator to tolerate faults on the Camera leads to a new potential threat.
This potential threat, however, does not lead to a threat since an intruder cannot reach the monitor from public elements.
The deployment of the dual self-checking pair pattern to tolerate faults on the Body Control leads to a new threat since the ECU \code{ecu2} (\ie, Body Control) reads from the CAN \code{can2}.
This threat can be, in principle, mitigated by the same \code{firewall} (illustrated in Figure~\ref{fig:hls-security-patterns}) deployed between \code{ecu3} and \code{can2}.

\subsubsection{Safety Consequences Caused by Security Patterns}
We specified reasoning principles for determining when a security pattern can cause consequences to safety.
The deployment of a security pattern may lead to new faults and failures, as the deployed security pattern can be faulty, \eg, a firewall might erroneously block messages.
The following reasoning rules specifies the consequences of deploying the firewall pattern.
There is a new fault \code{IDPAT} (same ID of the pattern) associated with the firewall \code{FW}.
This fault triggers a failure \code{IDFL} of type erroneous.   

\begin{flushleft}
\small{
\code{ft(IDPAT,[FW]) :- securityPattern(IDPAT,[firewall,[\_,\_,FW],\_,\_,\_])}.\\

\code{fl(IDFL,[err]) :- securityPattern(IDPAT,[firewall,[\_,\_,FW],\_,\_,\_]),}\\ 
\hspace{1cm}\code{ft(IDPAT,[FW]), createID(IDPAT,firewall,IDFL).}\\

\code{ft2fl(IDPAT,IDFL) :- securityPattern(IDPAT,[firewall,[\_,\_,FW],\_,\_,\_]),}\\
\hspace{1cm}\code{ft(IDPAT,[FW]), fl(IDFL,[err]), createID(IDPAT,firewall,IDFL).}
}
\end{flushleft}

\begin{wrapfigure}{r}{0.45\textwidth}
  \begin{center}
    \includegraphics[width=0.43\textwidth]{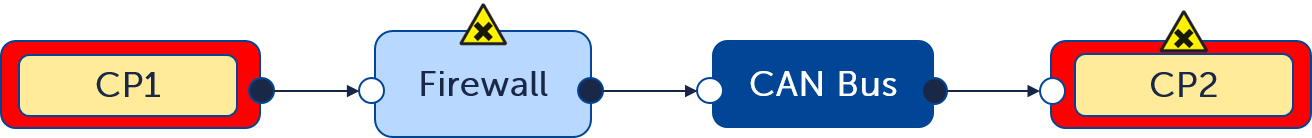}
  \end{center}
  \caption{Illustration of a cascading failure due to the deployment of a firewall}
  \label{fig:cascading-failure-fw}
\end{wrapfigure}

We also specify rules to check whether there is any cascading failure due to the deployment of a security pattern, \ie, to check whether the fault associated to a security pattern triggers an identified hazard.
Figure~\ref{fig:cascading-failure-fw} illustrates a cascading failure caused by the deployment of a firewall. 
The behavior of component \code{CP2} depends on the signals sent by component \code{CP1}, and there is fault associated to \code{CP2} that triggers failures leading to an identified hazard \code{HZ}.
If the deployed \code{Firewall} erroneously blocks messages from \code{CP1} then the failures on \code{CP1} might occur.
As a result, the failures from the \code{Firewall} might as well lead to hazard \code{HZ}.

\subsubsection{Safety Consequences on the Headlamp System}
Consider the headlamp system with the firewall pattern illustrated in Figure~\ref{fig:hls-security-patterns}.
The deployment of the firewall pattern leads to a new fault (triggering erroneous failures) on the firewall. 
These failures might lead to a cascading failure effecting the Body Control (\eg, when a user attempts to perform a software update via Bluetooth). 
However, since the functionality of the headlamp system is independent of its external components, this cascading failure might not cause any harm.

\subsubsection{Safety and Security Consequences}

An integration activity is required to harmonize the safety and security consequences.
This activity requires a manual analysis by safety and security engineers to assess the impact of the new faults and threats caused by the deployment of architecture patterns.

The manual analysis may include the assessment of whether (1) the new faults may lead to high criticality hazards and (2) the new threats are associated with components placed at communication channels with safety-relevant information.
In case scenarios (1) or (2) are found, safety and security engineers may either use measures like testing, simulation, or formal verification techniques to minimize the risks of these faults and threats or run \machinery to recommend further architecture patterns.
For example, \machinery may recommend the Heterogeneous Duplex pattern associated with a faulty firewall, where the second instance of the firewall shall be implemented by a different security team, and the fault detector shall check whether the outputs from both firewalls match.
Notice, however, that when using \machinery new safety and security consequences will be found. 
As a result, another manual analysis shall be carried out until a consensus is found between safety and security engineers.
We leave to future work the investigation on how to improve \machinery to find optional design solutions between safety and security consequences.

\section{Related Work}\label{sec:related-work}

Similar to our approach, \cite{DelmasDP15} proposes a methodology to harden system architectures by automating the choice of safety patterns to avoid failures. 
They provide a hardening strategy that consists of: (1) \emph{Component selection} that selects a component of the architecture that a safety pattern shall be added. (2) \emph{Pattern selection} that selects a pattern from a pre-defined library of patterns, and (3) \emph{Component substitution} that replaces the selected component by its hardened version with a safety pattern. 
This strategy is automated by the SAT4J solver~\cite{BerreP10}. 
The key difference to our work is that we provide means to harden system architectures with security patterns, in addition to safety patterns.
We also provide means to automate the consequences of applying security patterns to safety and vice versa.
Safety-wise, our reasoning principles enables a more precise recommendation of safety patterns as we specify a detailed intent for each safety pattern.

Once a safety pattern is selected, there shall be assumptions to ensure that the selected pattern is correctly applied to the system.
Recently, \cite{SljivoUPG20} proposed a methodology for assuring the application of safety and security patterns using contracts.
By \emph{contract}, they refer to a pair of assumptions and properties such that the properties only hold if the assumptions also hold.
Relying on the instantiation of a pattern template that includes both assumptions and properties (similar to the template used by this paper), they proposed a safety case argument pattern to guide the assurance of systems using patterns.
The specification of architecture pattern contracts are, however, done using informal descriptions only.
We provide a specification of architecture patterns that enables automation, including the generation of assumptions for each architecture pattern.
As future work, we plan to investigate how to extend \machinery to support architecture pattern with contracts.

Safety and security co-analysis using patterns has been addressed by some previous work~\cite{preschern13plop,MartinMSWKSAMK20}. 
We have been greatly inspired by~\cite{MartinMSWKSAMK20} that proposed a pattern-based approach for safety and security co-analysis, and by~\cite{preschern13plop} with security analysis of safety patterns. 
A key difference to our work is that we propose automated reasoning methods with safety and security patterns, whereas previous activities were done manually.

Model-based models and methods have been proposed for safety and security co-analysis, using languages such as GSN and Attack Trees and their combination~\cite{SadanySK19,kondeva19wosocer,preschern13plop,Pedroza18}. 
The key purpose of these approaches is to elucidate and document arguments demonstrating safety and security. 
Therefore, the artifacts produced often lie in high-levels of abstraction, \eg, expressing high-level safety and security goals or do not address the fact that safety and security have different semantics and different risk assessment methods.
Our work complements these approaches by providing means to automate reasoning based on declarative semantics provided by answer-set programs.
Indeed. we have developed a plugin that integrates the safety-related parts of \machinery into the model-based system engineering tool AutoFOCUS3~\cite{af3} to provide automated safety reasoning in a model-based engineering development~\cite{Dantas22modelsward}.

Some previous works proposed the use of security guide-words to identify information that is relevant for safety~\cite{durrwang18sae,durrwang17safecomp}.
For example,~\cite{durrwang18sae} provided a mapping involving SGM guide-words, CIA triad, and STRIDE nomenclature.
Using threat categories (\eg, STRIDE) enables a systematic identification of threat scenarios, possibly easing the recommendation of security patterns.
We believe that finer reasoning principles can be obtained by using more specific guide-words such as those proposed by~\cite{durrwang18sae,durrwang17safecomp}. 
This is left for future work.

Finally, we have recently demonstrated in white paper~\cite{dantas-whitepaper2020} that \machinery can also perform security analysis following the ISO 21434 risk assessment. 
\section{Conclusion}\label{sec:conclusion}
We have proposed the use of semantically-rich safety and security patterns for enabling automated support for safety and security co-design. 
In particular, we have proposed a Domain-Specific Language (DSL), called \machinery, that includes several safety and security concepts. 
We demonstrated its use in the description of several well-known patterns. 
Moreover, by using KRR methods, we demonstrate how one can specify reasoning principles as answer-set programs. 
As a result, it can automate several activities, \eg, when a potential threat can be derived from identified failures, and when a potential threat becomes a threat (including the attack path).
Moreover, \machinery can automatically recommend which pattern can be used at which place of the system architecture to address failures or threats, as well as make explicit consequences of deploying such patterns.

We are investigating how to extend our plugin~\cite{Dantas22modelsward} to integrate the security-related parts of \machinery into the model-based engineering tool AutoFOCUS3~\cite{af3}.
The scalability of \machinery shall also be investigated.
\machinery currently provides all possible architecture solutions to the user. 
In our recent paper~\cite{Dantas22modelsward}, we have defined four criteria to help the user in selecting the most suitable architecture for the system. 
We have carried out some initial experiments regarding the computation time of \machinery. 
We believe that the computation time of \machinery increases depending on the number of safety or security elements (\eg, on the number of faults). 
We have applied \machinery to an industrial use case taken from the automotive domain~\cite{Dantas22modelsward}, where eight faults have been identified.
\machinery took around ten minutes to compute all solutions. 
However, given that the focus of \machinery is on development time and not runtime, \machinery's performance requirements may range on hours or even days.
In the future, a dedicated study shall be carried out to determine exactly the scalability of \machinery.

\bibliographystyle{ACM-Reference-Format}
\bibliography{sample-base}

\end{document}